\documentclass[aps,prl,reprint,a4paper,floatfix,showpacs,showkeys,superscriptaddress]{revtex4-1}

\usepackage{graphicx}
\usepackage{amsmath}
\usepackage{epstopdf}

\newcommand{\mytilde}{\raise.17ex\hbox{$\scriptstyle\mathtt{\sim}$}}

\begin{document}

\title{Encounter-Limited Charge Carrier Recombination in Phase Separated Organic Semiconductor Blends}

\author{Michael C. Heiber}
\email{heiber@mailaps.org}
\affiliation{Experimental Physics VI, Julius-Maximilians-University of W{\"u}rzburg, 97074 W{\"u}rzburg, Germany}
\affiliation{Institut f{\"u}r Physik, Technische Universit{\"a}t Chemnitz, 09126 Chemnitz, Germany}

\author{Christoph Baumbach}
\affiliation{Institut f{\"u}r Physik, Technische Universit{\"a}t Chemnitz, 09126 Chemnitz, Germany}

\author{Vladimir Dyakonov}
\affiliation{Experimental Physics VI, Julius-Maximilians-University of W{\"u}rzburg, 97074 W{\"u}rzburg, Germany}%
\affiliation{Bavarian Centre for Applied Energy Research (ZAE Bayern), 97074 W{\"u}rzburg, Germany}

\author{Carsten Deibel}
\email{deibel@physik.tu-chemnitz.de}
\affiliation{Institut f{\"u}r Physik, Technische Universit{\"a}t Chemnitz, 09126 Chemnitz, Germany}

\date{\today}

\begin{abstract}
The theoretical effects of phase separation on encounter-limited charge carrier recombination in organic semiconductor blends are investigated using kinetic Monte Carlo (KMC) simulations of pump-probe experiments. Using model bulk heterojunction morphologies, the dependence of the recombination rate on domain size and charge carrier mobility are quantified. Unifying competing models and simulation results, we show that the mobility dependence of the recombination rate can be described using the power mean of the electron and hole mobilities with a domain size dependent exponent.  Additionally, for domain sizes typical of organic photovoltaic devices, we find that phase separation reduces the recombination rate by less than one order of magnitude compared to the Langevin model and that the mobility dependence can be approximated by the geometric mean. 
\end{abstract}



\maketitle


Despite vast scientific investigations on organic electronic devices over the last two decades, significant gaps in fundamental understanding exist in key areas.  A detailed description of charge carrier recombination, important for designing organic photovoltaics (OPVs), organic light-emitting diodes, and organic photodiodes, is a work in progress. To continue improving these devices, the dominant factors controlling recombination processes must be understood further so that they can be carefully controlled in energy efficient devices. 

Bimolecular charge recombination in organic semiconductors is most commonly described as a second-order reaction following the Langevin model,\cite{langevin1903}
\begin{equation}
	R = k_\text{L} np,
	\label{eq:rate}
\end{equation}
where $k_\text{L}$ is the Langevin recombination coefficient and $n$ and $p$ are the concentrations of electrons and holes, respectively. $k_\text{L}$ is derived by assuming an encounter-limited reaction in which the time it takes for an electron and hole to come together due to their Coulomb attraction is rate limiting.  As a result,
\begin{equation}
	k_\text{L} = \frac{e}{\epsilon\epsilon_0} (\mu_e + \mu_h),
	\label{eqn:langevin}
\end{equation}
where $e$ is the elementary charge, $\epsilon$ is the dielectric constant, $\epsilon_0$ is the vacuum permittivity, and $\mu_e$ and $\mu_h$ are the electron and hole mobilities, respectively.  

The Langevin model also assumes a spatially and energetically homogeneous and isotropic system with no internal electric field, which is not strictly valid in most organic semiconducting devices.  Organic semiconductors are well-characterized as having varying degrees of energetic and spatial disorder that can have a major impact on charge transport properties as highlighted by the commonly used Gaussian disorder model (GDM).\cite{bassler1993} In addition, devices may operate with a significant internal electric field. Investigating these issues, a number of kinetic Monte Carlo (KMC) simulation studies have identified conditions where deviations from the Langevin model occur.\cite{ries1984,richert1989,albrecht1995,gartstein1996,groves2008b}  
However, \citeauthor{vanderholst2009} concluded that the Langevin model still works well in an isotropic system with the GDM at low electric fields as long as accurate mobility values are used.\cite{vanderholst2009}  In agreement, measurements on a number of neat organic semiconducting materials have been consistent with the Langevin model.\cite{blom1997,pivrikas2005a,vanmensfoort2011,gorenflot2014} 

While the Langevin model may work well for neat materials, many devices utilize phase separated blends.\cite{mcneill2009} In such blends, electrons and holes are relegated to separate phases and are only able to undergo recombination at the phase boundaries. The resulting spatial limitations on charge carrier motion and recombination locations are expected to alter the recombination kinetics. Bulk heterojunction (BHJ) OPVs with domain sizes ranging from approximately 10 to 50~nm have often exhibited two major deviations from the Langevin model.  First, super-second order recombination kinetics has been measured in several blend systems.\cite{montanari2002,kim2006a,deibel2008,juska2008b,clarke2012a,rauh2012,foertig2013}
Often attributed to charge traps, the reasons for this behavior are still under debate.\cite{clarke2009,foertig2009,juska2009,shuttle2010a,kirchartz2012,rauh2012,deibel2013,gorenflot2014}

The second major deviation, commonly observed in P3HT:PCBM films, is a recombination rate that is several orders of magnitude less than predicted by the Langevin model.\cite{pivrikas2005b,juska2005,shuttle2008b,deibel2008} As a result, \citeauthor{pivrikas2005b} proposed that a reduced recombination rate is an inherent property of BHJ blends,\cite{pivrikas2005b} and \citeauthor{koster2006a} created the minimum mobility model, arguing that the recombination rate in a phase separated system should be limited by the mobility of the slowest carrier,\cite{koster2006a}
\begin{equation}
	k_\text{min} = \frac{e}{\epsilon\epsilon_0} \text{min}(\mu_e , \mu_h).
	\label{eqn:min_mobility}
\end{equation}
However, even though a significant reduction was also found in several other blends,\cite{scharber2010,clarke2011,murthy2013,albrecht2012,wetzelaer2013,roland2014}
some blend systems exhibit recombination rates much closer to the Langevin model.\cite{mozer2005c,dennler2006,scharber2010,clarke2011,clarke2012b}
It is becoming increasingly clear that a greatly reduced recombination rate is not an inherent property of phase separated blends but is a property that is dependent on a number of factors that are still under debate.\cite{juska2009,szmytkowski2009,deibel2009b,hilczer2010,howard2010b,ferguson2011,murthy2013,gorenflot2014}  

Despite clear deviations in many cases, the Langevin model is often used to explain bimolecular recombination in phase separated blends because a more complete model is still missing. An important first step towards a more complete model is understanding the fundamental effect of phase separation on encounter-limited bimolecular recombination.  Previously, \citeauthor{groves2008b} used KMC simulations to show that the recombination rate in a simple phase separated system lies somewhere between the Langevin model and the minimum mobility model with a weak dependence on domain size.\cite{groves2008b}  In addition, some experiments have also indicated a relatively weak domain size dependence.\cite{albrecht2012} However, other KMC simulations\cite{hamilton2010} and experiments\cite{roland2014} have indicated that the domain size could have a larger impact.  

In this letter, we determine the dependence of the recombination rate on the domain size and the electron and hole mobilities. We find that with a very small domain size, the Langevin model still holds, but for larger domains, clear deviations are present.  Unifying the competing Langevin and minimum mobility models with our simulation results, we show that the mobility dependence can be described using the power mean of the mobilities with a domain size dependent exponent.  Additionally, for domain sizes typical of OPVs, we find that phase separation reduces the recombination rate by less than one order of magnitude compared to the Langevin model and that the mobility dependence can be approximated by the geometric mean. 

Due to the complex geometry of phase separated systems, analytical derivation of the recombination rate as a function of the domain size is extremely difficult.  As a result, KMC simulations were performed to reach a numerical solution.  The simulations were configured to simulate pump-probe experiments on BHJ films without electrodes.  Using the Ising\_OPV software tool, a simple Ising phase separation model was implemented to create morphologies with varying domain size.\cite{heiber2014a,heiber2014c} 
To verify the presence of each recombination regime, the domain size was varied from 5 to 55~nm.  For each domain size ($d$) tested, 100 morphologies were independently generated to form a morphology set. The resulting morphologies consist of two pure phases with equivalent average domain sizes in a bicontinuous network configuration.

The model morphologies were then implemented into a three-dimensional lattice with a lattice constant of 1~nm, and sites were assigned energies from an uncorrelated Gaussian DOS with an energetic disorder ($\sigma$) of 75~meV. Two-dimensional periodic boundary conditions were used to simulate a thin film.  To start the simulation, excitons were created with uniform probability throughout the lattice with a Gaussian excitation pulse having a pulse width of 100~ps and an intensity corresponding to an initial exciton concentration of $5 \times 10^{17}$~cm$^{-3}$.  However, for 55~nm domains, the initial exciton concentration was set to $3.1 \times 10^{17}$~cm$^{-3}$ due to computational limitations.  This change had no impact on the recombination behavior of interest in this study. The complexities of charge separation were bypassed to create free charge carriers directly from excitons. To do this, electron-hole pairs were created across the interface with a separation distance of 30~nm by restricting exciton creation to within 30~nm of an interface and executing an ultrafast long-range charge transfer event.
The charges in the lattice then underwent hopping transport using the Miller-Abrahams (MA) model,\cite{miller1960}
and coulomb interactions were included between charges within a cutoff radius of 35~nm.  Electron hopping was restricted to acceptor domains and hole hopping was restricted to donor domains. Charge recombination was also implemented using the MA model with a recombination prefactor ($R_\text{0,rec}$) that was held constant at a large value of $10^{15}$~s$^{-1}$ to ensure that recombination dominated over redissociation. 

For each simulation, 24 morphologies were randomly selected from the appropriate morphology set and 4 random configurations of energetic disorder were implemented for each, and the results of the 96 runs were averaged. During each simulation, the hole concentration was logged as a function of simulated time. With the lattice sizes used, the carrier concentration could be resolved over two and half orders of magnitude, covering a range typical for steady state illumination intensities from 0.1 to 10 suns. Assuming second-order recombination kinetics and $n=p$, the numerical derivative of the hole concentration as a function of time ($t$) was then used to calculate the simulated time-dependent recombination coefficient:
\begin{equation}
k_\text{sim}(t) = -\frac{\frac{dp(t)}{dt}}{p(t)^2}.
\label{eqn:k_sim}
\end{equation}
In addition, the displacement of each carrier from its initial position was recorded over its lifetime, and the numerical derivative of the average mean squared displacement over time for all carriers was used to calculate the average time-dependent diffusion coefficient. Due to the thin film geometry, the two-dimensional diffusion equation was found to be most appropriate. Using the Einstein relation, which remains valid at zero-field\cite{richert1989,nenashev2010b} when recombination removes deeply trapped carriers,\cite{wetzelaer2011b} the average time-dependent zero-field mobility of each carrier type was determined,
\begin{equation}
\mu(t) = \frac{e}{4 k_\text{B} T} \frac{d\langle r(t)^2 \rangle}{dt}.
\label{eqn:mobility}
\end{equation}
More information about the morphology generation, KMC simulation parameters, and data analysis is shown in the Supplementary Information.\footnotemark[1]

\begin{figure}[h]
\includegraphics[scale=1]{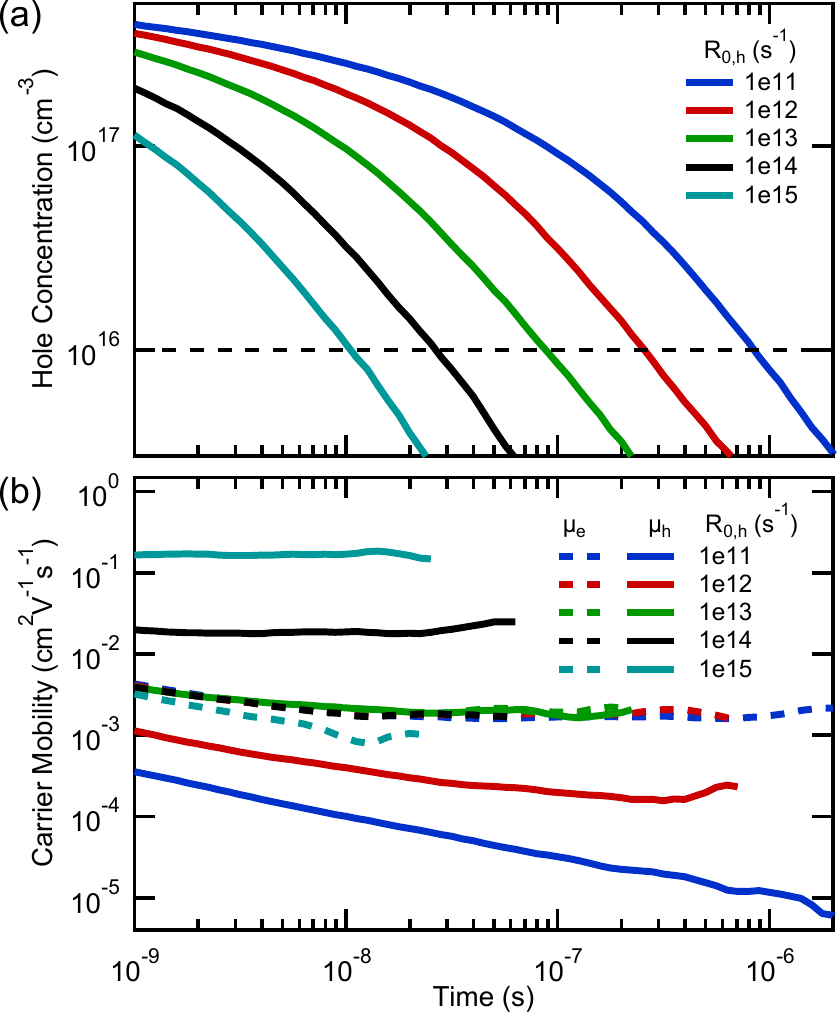}
\caption{\label{fig:transients}(a) Hole concentration transients and (b) time-dependent mobilities for $d=15$~nm with varying hole hopping rates.}
\end{figure}

To determine the mobility dependence, the relative magnitudes of the electron and hole mobilities were tuned by varying the hole hopping prefactor ($R_\text{0,h}$) from $10^{11}$ to $10^{15}$~s$^{-1}$ while holding the electron hopping prefactor ($R_\text{0,e}$) constant at $10^{13}$~s$^{-1}$. Figure \ref{fig:transients} shows how the hole concentration ($p$) decays over time for differing magnitudes of $R_\text{0,h}$. By the time the hole concentration reached $10^{16}$~cm$^{-3}$ (a typical concentration for 1 sun steady state illumination), all transients showed steady second-order decay.  The time point where $p=10^{16}$~cm$^{-3}$ was used as the comparison point between different simulations. At this time point, the simulated recombination coefficient ($k_\text{sim}$), the mobilities ($\mu_e$ and $\mu_h$), the Langevin recombination coefficient ($k_\text{L}$), and the minimum mobility model recombination coefficient ($k_\text{min}$) were calculated.  

\begin{figure}[h]
\includegraphics[scale=1]{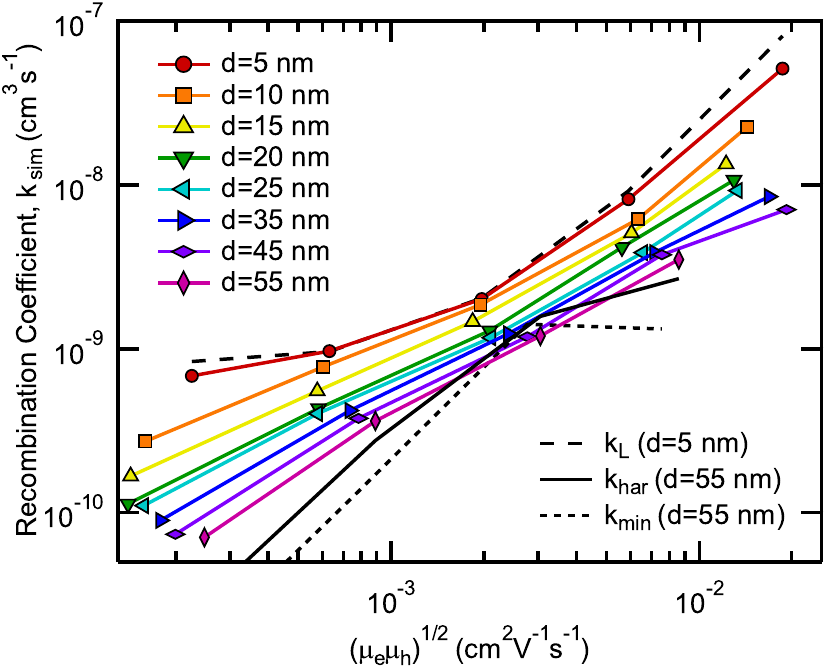}
\caption{\label{fig:main_results}The effect of the domain size on the mobility dependence of the simulated recombination rate coefficient compared to the Langevin, harmonic mean, and minimum mobility models.}
\end{figure}

Figure \ref{fig:main_results} shows how the mobility dependence of the recombination coefficient changes with different magnitudes of domain size. With a very small domain size of 5~nm, the recombination rate approaches the Langevin model, and as the domain size increases, the recombination rate deviates from the Langevin model.  In the intermediate regime, in agreement with the results of \citeauthor{groves2008b},\cite{groves2008b} the simulated recombination rate was not found to be proportional to the sum of mobilities nor the minimum mobility.  Instead, a new trend not captured by any of the current models is observed, in which $k_\text{sim}$ is approximately proportional to the geometric mean of the electron and hole mobilities.  For the largest domain sizes tested, the mobility dependence continues to change as observed by the downward curvature beginning to form for d=45 and 55~nm.  Here, the mobility dependence appears to be approaching the shape predicted by the minimum mobility model. However, the bias toward the minimum mobility appears much weaker than predicted in the minimum mobility model.  Another type of mean which favors the smaller value is the harmonic mean, and the simulated recombination behavior appears to be closer to the harmonic mean.  The same general behavior was also observed when using different magnitudes of energetic disorder ($\sigma$) and when performing the analysis at carrier concentrations from $5\times10^{15}$ to $2\times10^{16}$~cm$^{-3}$.  

\begin{figure}[ht]
\includegraphics[scale=1]{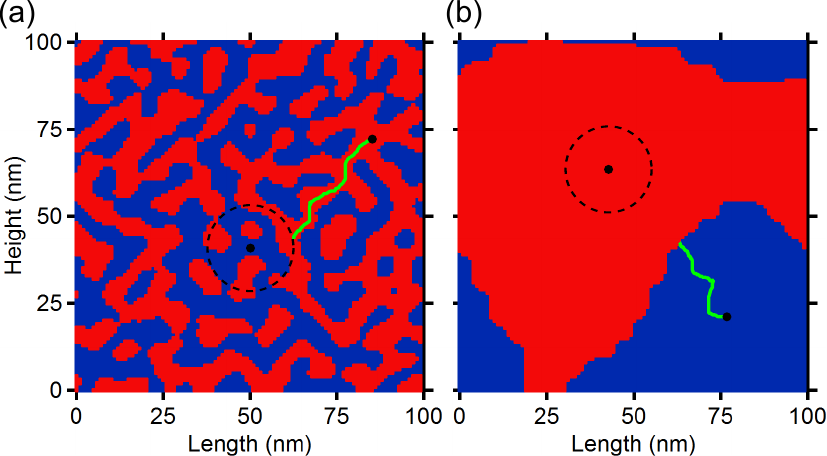}
\caption{\label{fig:morphologies}Depiction of the two extreme recombination regimes: (a) very small domain size and (b) very large domain size.  One charge is depicted with a dotted circle around it that represents the Coulomb capture radius ($R_C$).  The other charge is shown with a green path that illustrates one of the most likely pathways to enter the capture sphere.}
\end{figure}

To understand this behavior, consider a case where the electron has a very fast mobility compared to the hole.  In a neat material, even if the hole is slow, it is still accessible to the electron everywhere.  In this simple case, the resulting recombination rate is proportional to the sum of the individual mobilities as described in the Langevin model.  Now, let us consider a phase separated morphology with very small domains, as shown in Fig.\ \ref{fig:morphologies}a.  In this case, with highly interconnected domains and high interfacial area, the slow hole is still accessible to the electron almost everywhere.  When the electron enters the Coulomb capture radius of the hole (depicted by the dashed circle), the attractive force becomes very strong and recombination occurs quickly.  As a result, even though the electron and hole are restricted to separate phases, the resulting recombination rate should still follow the Langevin model.  

However, with very large domains, as shown in Fig.\ \ref{fig:morphologies}b, most of the slow holes are inaccessible to the electrons, and in order for recombination to occur, the hole must first migrate to the interface.  As a result, the recombination rate should be limited by the mobility of the slower hole and should follow the minimum mobility model.  Given these two well-defined extreme cases, it follows logically that for intermediate domain sizes, the recombination behavior should be somewhere inbetween.  In effect, the recombination rate of the slower carriers depends mainly on their distance from the interface.  Those located right next to the interface will require almost no further motion, and their recombination rate will be dominated by the speed of the faster carrier.  However, those located farther away will start to be limited more by their own slow motion, passing through a regime where the recombination rate will depend on the speed of both carriers and then eventually dominated by the slow carrier when very far away.  Given this behavior, when changing the domain size, the critical change is in the average distance between the carriers and the interface, which then alters the mobility dependence.

Capturing this behavior and unifying the competing analytical models with the simulations results of \citeauthor{groves2008b}\cite{groves2008b} and those presented here, all curves in Fig.\ \ref{fig:main_results} can be fit by the following simple equation (see Supplementary Information),\footnotemark[1]
\begin{equation}
k_\text{sim} =  \frac{e}{\epsilon\epsilon_0} f_1(d) 2 M_{g(d)}(\mu_e,\mu_h),
\end{equation}
where $f_1(d)$ is a domain size dependent prefactor and $M_{g(d)}(\mu_e,\mu_h)$ is the power mean (generalized mean),
\begin{equation}
M_g(\mu_e,\mu_h) = \left(\frac{\mu_e^g+\mu_h^g}{2}\right)^{1/g},
\end{equation}
with a domain size dependent exponent, $g(d)$.
For a very small domain size, $\lim_{d \to 0} f_1(d) = 1$ and $\lim_{d \to 0} g(d) = 1$, and the Langevin expression is obtained.  In addition, $\lim_{g \to 0} M_g$ is the geometric mean, $M_{-1}$ is the harmonic mean, and  $M_{-\infty}$ is the minimum value.  

\begin{figure}[h]
\includegraphics[scale=1]{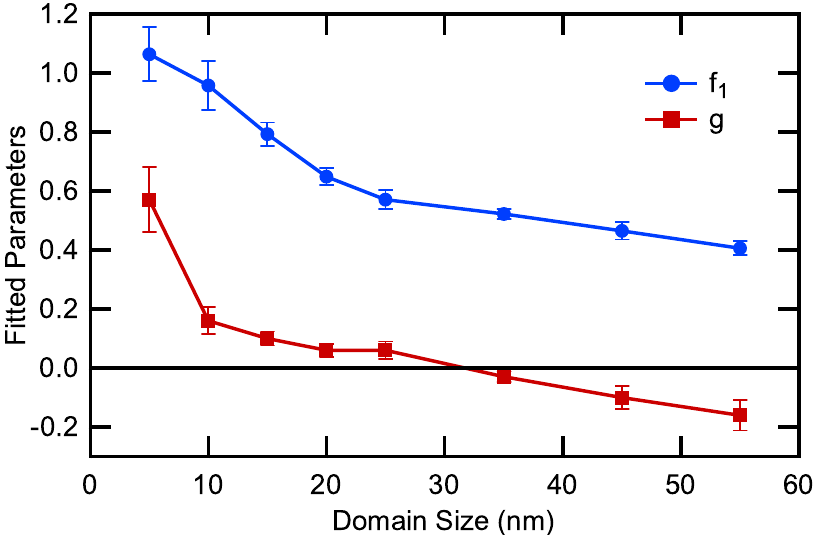}
\caption{\label{fig:fit_params}Fitted values for the prefactor ($f_1$) and the power mean exponent ($g$) as a function of the domain size.}
\end{figure}

Fig.\ \ref{fig:fit_params} shows the fitted values for $f_1$ and $g$ as a function of the domain size.  For $d=5$~nm, $f_1$ and $g$ both approach one, meaning that the behavior is very close to the Langevin model.  For intermediate domain sizes of about 10-35~nm, $g\approx 0$, meaning that the behavior in this regime can be approximated by the geometric mean.  Then for $d>35$~nm, $g$ continues to slowly decrease, thereby increasing the impact of the minimum mobility.  

In addition to the change in the mobility dependence, the magnitude of the recombination rate also decreases with increasing domain size.  It is clear that simple phase separation alone causes only a mild reduction of less than one order of magnitude when the electron and hole mobilities are within one order of magnitude of each other, as is the case in most optimized devices.  As a result, blends with much larger reductions must have other more dominant contributing factors. These blends are likely to be greatly affected by polaron pair redissociation and are therefore not in the encounter-limited regime and/or are affected by interfacial states different from those in the bulk.\cite{gorenflot2014} 


In conclusion, we have quantified the effect of phase separation on the bimolecular charge recombination rate when in the encounter-limited regime.  Most significantly, we have shown that the mobility dependence can be described by the power mean with a domain size dependent exponent.  For domain sizes typical of OPVs, the geometric mean is a very good approximation of the mobility dependence.  In addition, we clearly demonstrate that greatly reduced recombination rates are not an inherent property of phase separated systems.  These results represent a unification of previous recombination models and a major step forward in developing a more complete model for charge carrier recombination in organic semiconductor blends.  With this knowledge, updated design rules for new materials and device architectures can be defined.


\begin{acknowledgments}
M.C.H.\ and C.D.\ acknowledge funding by Deutsche Forschungsgemeinschaft (DFG) grant DE830/13-1. C.D.\ thanks Prof. Michael Schreiber for a brief but helpful discussion.  We also thank Dr.\ Kristofer Tvingstedt and Dr.\ Andreas Baumann for insightful discussion and feedback on the manuscript.
\end{acknowledgments}

\footnotetext[1]{See Supplemental Material at [URL] for morphology generation details, a full list of simulation parameters, and supplementary results.}





\begin{thebibliography}{3}%
\makeatletter
\providecommand \@ifxundefined [1]{%
 \@ifx{#1\undefined}
}%
\providecommand \@ifnum [1]{%
 \ifnum #1\expandafter \@firstoftwo
 \else \expandafter \@secondoftwo
 \fi
}%
\providecommand \@ifx [1]{%
 \ifx #1\expandafter \@firstoftwo
 \else \expandafter \@secondoftwo
 \fi
}%
\providecommand \natexlab [1]{#1}%
\providecommand \enquote  [1]{``#1''}%
\providecommand \bibnamefont  [1]{#1}%
\providecommand \bibfnamefont [1]{#1}%
\providecommand \citenamefont [1]{#1}%
\providecommand \href@noop [0]{\@secondoftwo}%
\providecommand \href [0]{\begingroup \@sanitize@url \@href}%
\providecommand \@href[1]{\@@startlink{#1}\@@href}%
\providecommand \@@href[1]{\endgroup#1\@@endlink}%
\providecommand \@sanitize@url [0]{\catcode `\\12\catcode `\$12\catcode
  `\&12\catcode `\#12\catcode `\^12\catcode `\_12\catcode `\%12\relax}%
\providecommand \@@startlink[1]{}%
\providecommand \@@endlink[0]{}%
\providecommand \url  [0]{\begingroup\@sanitize@url \@url }%
\providecommand \@url [1]{\endgroup\@href {#1}{\urlprefix }}%
\providecommand \urlprefix  [0]{URL }%
\providecommand \Eprint [0]{\href }%
\providecommand \doibase [0]{http://dx.doi.org/}%
\providecommand \selectlanguage [0]{\@gobble}%
\providecommand \bibinfo  [0]{\@secondoftwo}%
\providecommand \bibfield  [0]{\@secondoftwo}%
\providecommand \translation [1]{[#1]}%
\providecommand \BibitemOpen [0]{}%
\providecommand \bibitemStop [0]{}%
\providecommand \bibitemNoStop [0]{.\EOS\space}%
\providecommand \EOS [0]{\spacefactor3000\relax}%
\providecommand \BibitemShut  [1]{\csname bibitem#1\endcsname}%
\let\auto@bib@innerbib\@empty
\bibitem [{\citenamefont {Heiber}(2014)}]{heiber2014a}%
  \BibitemOpen
  \bibfield  {author} {\bibinfo {author} {\bibfnamefont {M.~C.}\ \bibnamefont
  {Heiber}},\ }\href@noop {} {\enquote {\bibinfo {title} {{I}sing\_{OPV}},}\
  }\bibinfo {howpublished} {\url{https://github.com/MikeHeiber/Ising_OPV}}
  (\bibinfo {year} {2014})\BibitemShut {NoStop}%
\bibitem [{\citenamefont {Heiber}\ and\ \citenamefont
  {Dhinojwala}(2014)}]{heiber2014c}%
  \BibitemOpen
  \bibfield  {author} {\bibinfo {author} {\bibfnamefont {M.~C.}\ \bibnamefont
  {Heiber}}\ and\ \bibinfo {author} {\bibfnamefont {A.}~\bibnamefont
  {Dhinojwala}},\ }\href {\doibase 10.1103/PhysRevApplied.2.014008} {\bibfield
  {journal} {\bibinfo  {journal} {Phys. Rev. Appl.}\ }\textbf {\bibinfo
  {volume} {2}},\ \bibinfo {pages} {014008} (\bibinfo {year}
  {2014})}\BibitemShut {NoStop}%
\bibitem [{\citenamefont {Heiber}\ and\ \citenamefont
  {Dhinojwala}(2012)}]{heiber2012}%
  \BibitemOpen
  \bibfield  {author} {\bibinfo {author} {\bibfnamefont {M.~C.}\ \bibnamefont
  {Heiber}}\ and\ \bibinfo {author} {\bibfnamefont {A.}~\bibnamefont
  {Dhinojwala}},\ }\href {\doibase 10.1063/1.4731698} {\bibfield  {journal}
  {\bibinfo  {journal} {J. Chem. Phys.}\ }\textbf {\bibinfo {volume} {137}},\
  \bibinfo {pages} {014903} (\bibinfo {year} {2012})}\BibitemShut {NoStop}%
\end{thebibliography}%


\begin{thebibliography}{51}%
\makeatletter
\providecommand \@ifxundefined [1]{%
 \@ifx{#1\undefined}
}%
\providecommand \@ifnum [1]{%
 \ifnum #1\expandafter \@firstoftwo
 \else \expandafter \@secondoftwo
 \fi
}%
\providecommand \@ifx [1]{%
 \ifx #1\expandafter \@firstoftwo
 \else \expandafter \@secondoftwo
 \fi
}%
\providecommand \natexlab [1]{#1}%
\providecommand \enquote  [1]{``#1''}%
\providecommand \bibnamefont  [1]{#1}%
\providecommand \bibfnamefont [1]{#1}%
\providecommand \citenamefont [1]{#1}%
\providecommand \href@noop [0]{\@secondoftwo}%
\providecommand \href [0]{\begingroup \@sanitize@url \@href}%
\providecommand \@href[1]{\@@startlink{#1}\@@href}%
\providecommand \@@href[1]{\endgroup#1\@@endlink}%
\providecommand \@sanitize@url [0]{\catcode `\\12\catcode `\$12\catcode
  `\&12\catcode `\#12\catcode `\^12\catcode `\_12\catcode `\%12\relax}%
\providecommand \@@startlink[1]{}%
\providecommand \@@endlink[0]{}%
\providecommand \url  [0]{\begingroup\@sanitize@url \@url }%
\providecommand \@url [1]{\endgroup\@href {#1}{\urlprefix }}%
\providecommand \urlprefix  [0]{URL }%
\providecommand \Eprint [0]{\href }%
\providecommand \doibase [0]{http://dx.doi.org/}%
\providecommand \selectlanguage [0]{\@gobble}%
\providecommand \bibinfo  [0]{\@secondoftwo}%
\providecommand \bibfield  [0]{\@secondoftwo}%
\providecommand \translation [1]{[#1]}%
\providecommand \BibitemOpen [0]{}%
\providecommand \bibitemStop [0]{}%
\providecommand \bibitemNoStop [0]{.\EOS\space}%
\providecommand \EOS [0]{\spacefactor3000\relax}%
\providecommand \BibitemShut  [1]{\csname bibitem#1\endcsname}%
\let\auto@bib@innerbib\@empty
\bibitem [{\citenamefont {Langevin}(1903)}]{langevin1903}%
  \BibitemOpen
  \bibfield  {author} {\bibinfo {author} {\bibfnamefont {P.}~\bibnamefont
  {Langevin}},\ }\href@noop {} {\bibfield  {journal} {\bibinfo  {journal} {Ann.
  Chim. Phys.}\ }\textbf {\bibinfo {volume} {28}},\ \bibinfo {pages} {433}
  (\bibinfo {year} {1903})}\BibitemShut {NoStop}%
\bibitem [{\citenamefont {B{\"{a}}ssler}(1993)}]{bassler1993}%
  \BibitemOpen
  \bibfield  {author} {\bibinfo {author} {\bibfnamefont {H.}~\bibnamefont
  {B{\"{a}}ssler}},\ }\href {\doibase 10.1002/pssb.2221750102} {\bibfield
  {journal} {\bibinfo  {journal} {Phys. Status Solidi B}\ }\textbf {\bibinfo
  {volume} {175}},\ \bibinfo {pages} {15} (\bibinfo {year} {1993})}\BibitemShut
  {NoStop}%
\bibitem [{\citenamefont {Ries}\ and\ \citenamefont
  {B{\"a}ssler}(1984)}]{ries1984}%
  \BibitemOpen
  \bibfield  {author} {\bibinfo {author} {\bibfnamefont {B.}~\bibnamefont
  {Ries}}\ and\ \bibinfo {author} {\bibfnamefont {H.}~\bibnamefont
  {B{\"a}ssler}},\ }\href {\doibase 10.1016/0009-2614(84)80369-3} {\bibfield
  {journal} {\bibinfo  {journal} {Chem. Phys. Lett.}\ }\textbf {\bibinfo
  {volume} {108}},\ \bibinfo {pages} {71} (\bibinfo {year} {1984})}\BibitemShut
  {NoStop}%
\bibitem [{\citenamefont {Richert}\ \emph {et~al.}(1989)\citenamefont
  {Richert}, \citenamefont {Pautmeier},\ and\ \citenamefont
  {B{\"a}ssler}}]{richert1989}%
  \BibitemOpen
  \bibfield  {author} {\bibinfo {author} {\bibfnamefont {R.}~\bibnamefont
  {Richert}}, \bibinfo {author} {\bibfnamefont {L.}~\bibnamefont {Pautmeier}},
  \ and\ \bibinfo {author} {\bibfnamefont {H.}~\bibnamefont {B{\"a}ssler}},\
  }\href {\doibase 10.1103/PhysRevLett.63.547} {\bibfield  {journal} {\bibinfo
  {journal} {Phys. Rev. Lett.}\ }\textbf {\bibinfo {volume} {63}},\ \bibinfo
  {pages} {547} (\bibinfo {year} {1989})}\BibitemShut {NoStop}%
\bibitem [{\citenamefont {Albrecht}\ and\ \citenamefont
  {B{\"a}ssler}(1995)}]{albrecht1995}%
  \BibitemOpen
  \bibfield  {author} {\bibinfo {author} {\bibfnamefont {U.}~\bibnamefont
  {Albrecht}}\ and\ \bibinfo {author} {\bibfnamefont {H.}~\bibnamefont
  {B{\"a}ssler}},\ }\href {\doibase 10.1016/0009-2614(95)00121-J} {\bibfield
  {journal} {\bibinfo  {journal} {Chem. Phys. Lett.}\ }\textbf {\bibinfo
  {volume} {235}},\ \bibinfo {pages} {389} (\bibinfo {year}
  {1995})}\BibitemShut {NoStop}%
\bibitem [{\citenamefont {Gartstein}\ \emph {et~al.}(1996)\citenamefont
  {Gartstein}, \citenamefont {Conwell},\ and\ \citenamefont
  {Rice}}]{gartstein1996}%
  \BibitemOpen
  \bibfield  {author} {\bibinfo {author} {\bibfnamefont {Y.~N.}\ \bibnamefont
  {Gartstein}}, \bibinfo {author} {\bibfnamefont {E.~M.}\ \bibnamefont
  {Conwell}}, \ and\ \bibinfo {author} {\bibfnamefont {M.~J.}\ \bibnamefont
  {Rice}},\ }\href {\doibase 10.1016/0009-2614(95)01435-7} {\bibfield
  {journal} {\bibinfo  {journal} {Chem. Phys. Lett.}\ }\textbf {\bibinfo
  {volume} {249}},\ \bibinfo {pages} {451} (\bibinfo {year}
  {1996})}\BibitemShut {NoStop}%
\bibitem [{\citenamefont {Groves}\ and\ \citenamefont
  {Greenham}(2008)}]{groves2008b}%
  \BibitemOpen
  \bibfield  {author} {\bibinfo {author} {\bibfnamefont {C.}~\bibnamefont
  {Groves}}\ and\ \bibinfo {author} {\bibfnamefont {N.~C.}\ \bibnamefont
  {Greenham}},\ }\href {\doibase 10.1103/PhysRevB.78.155205} {\bibfield
  {journal} {\bibinfo  {journal} {Phys. Rev. B}\ }\textbf {\bibinfo {volume}
  {78}},\ \bibinfo {pages} {155205} (\bibinfo {year} {2008})}\BibitemShut
  {NoStop}%
\bibitem [{\citenamefont {van~der Holst}\ \emph {et~al.}(2009)\citenamefont
  {van~der Holst}, \citenamefont {van Oost}, \citenamefont {Coehoorn},\ and\
  \citenamefont {Bobbert}}]{vanderholst2009}%
  \BibitemOpen
  \bibfield  {author} {\bibinfo {author} {\bibfnamefont {J.~J.~M.}\
  \bibnamefont {van~der Holst}}, \bibinfo {author} {\bibfnamefont {F.~W.~A.}\
  \bibnamefont {van Oost}}, \bibinfo {author} {\bibfnamefont {R.}~\bibnamefont
  {Coehoorn}}, \ and\ \bibinfo {author} {\bibfnamefont {P.~A.}\ \bibnamefont
  {Bobbert}},\ }\href {\doibase 10.1103/PhysRevB.80.235202} {\bibfield
  {journal} {\bibinfo  {journal} {Phys. Rev. B}\ }\textbf {\bibinfo {volume}
  {80}},\ \bibinfo {pages} {235202} (\bibinfo {year} {2009})}\BibitemShut
  {NoStop}%
\bibitem [{\citenamefont {Blom}\ \emph {et~al.}(1997)\citenamefont {Blom},
  \citenamefont {de~Jong},\ and\ \citenamefont {Breedijk}}]{blom1997}%
  \BibitemOpen
  \bibfield  {author} {\bibinfo {author} {\bibfnamefont {P.~W.~M.}\
  \bibnamefont {Blom}}, \bibinfo {author} {\bibfnamefont {M.~J.~M.}\
  \bibnamefont {de~Jong}}, \ and\ \bibinfo {author} {\bibfnamefont
  {S.}~\bibnamefont {Breedijk}},\ }\href {\doibase 10.1063/1.119692} {\bibfield
   {journal} {\bibinfo  {journal} {Appl. Phys. Lett.}\ }\textbf {\bibinfo
  {volume} {71}},\ \bibinfo {pages} {930} (\bibinfo {year} {1997})}\BibitemShut
  {NoStop}%
\bibitem [{\citenamefont {Pivrikas}\ \emph
  {et~al.}(2005{\natexlab{a}})\citenamefont {Pivrikas}, \citenamefont
  {Ju{\v{s}}ka}, \citenamefont {{\"{O}}sterbacka}, \citenamefont {Westerling},
  \citenamefont {Vili{\={u}}nas}, \citenamefont {Arlauskas},\ and\
  \citenamefont {Stubb}}]{pivrikas2005a}%
  \BibitemOpen
  \bibfield  {author} {\bibinfo {author} {\bibfnamefont {A.}~\bibnamefont
  {Pivrikas}}, \bibinfo {author} {\bibfnamefont {G.}~\bibnamefont
  {Ju{\v{s}}ka}}, \bibinfo {author} {\bibfnamefont {R.}~\bibnamefont
  {{\"{O}}sterbacka}}, \bibinfo {author} {\bibfnamefont {M.}~\bibnamefont
  {Westerling}}, \bibinfo {author} {\bibfnamefont {M.}~\bibnamefont
  {Vili{\={u}}nas}}, \bibinfo {author} {\bibfnamefont {K.}~\bibnamefont
  {Arlauskas}}, \ and\ \bibinfo {author} {\bibfnamefont {H.}~\bibnamefont
  {Stubb}},\ }\href {\doibase 10.1103/PhysRevB.71.125205} {\bibfield  {journal}
  {\bibinfo  {journal} {Phys. Rev. B}\ }\textbf {\bibinfo {volume} {71}},\
  \bibinfo {pages} {125205} (\bibinfo {year} {2005}{\natexlab{a}})}\BibitemShut
  {NoStop}%
\bibitem [{\citenamefont {van Mensfoort}\ \emph {et~al.}(2011)\citenamefont
  {van Mensfoort}, \citenamefont {Billen}, \citenamefont {Carvelli},
  \citenamefont {Vulto}, \citenamefont {Janssen},\ and\ \citenamefont
  {Coehoorn}}]{vanmensfoort2011}%
  \BibitemOpen
  \bibfield  {author} {\bibinfo {author} {\bibfnamefont {S.~L.~M.}\
  \bibnamefont {van Mensfoort}}, \bibinfo {author} {\bibfnamefont
  {J.}~\bibnamefont {Billen}}, \bibinfo {author} {\bibfnamefont
  {M.}~\bibnamefont {Carvelli}}, \bibinfo {author} {\bibfnamefont {S.~I.~E.}\
  \bibnamefont {Vulto}}, \bibinfo {author} {\bibfnamefont {R.~A.~J.}\
  \bibnamefont {Janssen}}, \ and\ \bibinfo {author} {\bibfnamefont
  {R.}~\bibnamefont {Coehoorn}},\ }\href {\doibase 10.1063/1.3553412}
  {\bibfield  {journal} {\bibinfo  {journal} {J. Appl. Phys.}\ }\textbf
  {\bibinfo {volume} {109}},\ \bibinfo {pages} {064502} (\bibinfo {year}
  {2011})}\BibitemShut {NoStop}%
\bibitem [{\citenamefont {Gorenflot}\ \emph {et~al.}(2014)\citenamefont
  {Gorenflot}, \citenamefont {Heiber}, \citenamefont {Baumann}, \citenamefont
  {Lorrmann}, \citenamefont {Gunz}, \citenamefont {K{\"a}mpgen}, \citenamefont
  {Dyakonov},\ and\ \citenamefont {Deibel}}]{gorenflot2014}%
  \BibitemOpen
  \bibfield  {author} {\bibinfo {author} {\bibfnamefont {J.}~\bibnamefont
  {Gorenflot}}, \bibinfo {author} {\bibfnamefont {M.~C.}\ \bibnamefont
  {Heiber}}, \bibinfo {author} {\bibfnamefont {A.}~\bibnamefont {Baumann}},
  \bibinfo {author} {\bibfnamefont {J.}~\bibnamefont {Lorrmann}}, \bibinfo
  {author} {\bibfnamefont {M.}~\bibnamefont {Gunz}}, \bibinfo {author}
  {\bibfnamefont {A.}~\bibnamefont {K{\"a}mpgen}}, \bibinfo {author}
  {\bibfnamefont {V.}~\bibnamefont {Dyakonov}}, \ and\ \bibinfo {author}
  {\bibfnamefont {C.}~\bibnamefont {Deibel}},\ }\href {\doibase
  10.1063/1.4870805} {\bibfield  {journal} {\bibinfo  {journal} {J. Appl.
  Phys.}\ }\textbf {\bibinfo {volume} {115}},\ \bibinfo {pages} {144502}
  (\bibinfo {year} {2014})}\BibitemShut {NoStop}%
\bibitem [{\citenamefont {McNeill}\ and\ \citenamefont
  {Greenham}(2009)}]{mcneill2009}%
  \BibitemOpen
  \bibfield  {author} {\bibinfo {author} {\bibfnamefont {C.~R.}\ \bibnamefont
  {McNeill}}\ and\ \bibinfo {author} {\bibfnamefont {N.~C.}\ \bibnamefont
  {Greenham}},\ }\href {\doibase 10.1002/adma.200900783} {\bibfield  {journal}
  {\bibinfo  {journal} {Adv. Mater.}\ }\textbf {\bibinfo {volume} {21}},\
  \bibinfo {pages} {3840} (\bibinfo {year} {2009})}\BibitemShut {NoStop}%
\bibitem [{\citenamefont {Montanari}\ \emph {et~al.}(2002)\citenamefont
  {Montanari}, \citenamefont {Nogueira}, \citenamefont {Nelson}, \citenamefont
  {Durrant}, \citenamefont {Winder}, \citenamefont {Loi}, \citenamefont
  {Sariciftci},\ and\ \citenamefont {Brabec}}]{montanari2002}%
  \BibitemOpen
  \bibfield  {author} {\bibinfo {author} {\bibfnamefont {I.}~\bibnamefont
  {Montanari}}, \bibinfo {author} {\bibfnamefont {A.~F.}\ \bibnamefont
  {Nogueira}}, \bibinfo {author} {\bibfnamefont {J.}~\bibnamefont {Nelson}},
  \bibinfo {author} {\bibfnamefont {J.~R.}\ \bibnamefont {Durrant}}, \bibinfo
  {author} {\bibfnamefont {C.}~\bibnamefont {Winder}}, \bibinfo {author}
  {\bibfnamefont {M.~A.}\ \bibnamefont {Loi}}, \bibinfo {author} {\bibfnamefont
  {N.~S.}\ \bibnamefont {Sariciftci}}, \ and\ \bibinfo {author} {\bibfnamefont
  {C.}~\bibnamefont {Brabec}},\ }\href {\doibase 10.1063/1.1512943} {\bibfield
  {journal} {\bibinfo  {journal} {Appl. Phys. Lett.}\ }\textbf {\bibinfo
  {volume} {81}},\ \bibinfo {pages} {3001} (\bibinfo {year}
  {2002})}\BibitemShut {NoStop}%
\bibitem [{\citenamefont {Kim}\ \emph {et~al.}(2006)\citenamefont {Kim},
  \citenamefont {Cook}, \citenamefont {Tuladhar}, \citenamefont {Choulis},
  \citenamefont {Nelson}, \citenamefont {Durrant}, \citenamefont {Bradley},
  \citenamefont {Giles}, \citenamefont {McCulloch}, \citenamefont {Ha},\ and\
  \citenamefont {Ree}}]{kim2006a}%
  \BibitemOpen
  \bibfield  {author} {\bibinfo {author} {\bibfnamefont {Y.}~\bibnamefont
  {Kim}}, \bibinfo {author} {\bibfnamefont {S.}~\bibnamefont {Cook}}, \bibinfo
  {author} {\bibfnamefont {S.~M.}\ \bibnamefont {Tuladhar}}, \bibinfo {author}
  {\bibfnamefont {S.~A.}\ \bibnamefont {Choulis}}, \bibinfo {author}
  {\bibfnamefont {J.}~\bibnamefont {Nelson}}, \bibinfo {author} {\bibfnamefont
  {J.~R.}\ \bibnamefont {Durrant}}, \bibinfo {author} {\bibfnamefont
  {D.~D.~C.}\ \bibnamefont {Bradley}}, \bibinfo {author} {\bibfnamefont
  {M.}~\bibnamefont {Giles}}, \bibinfo {author} {\bibfnamefont
  {I.}~\bibnamefont {McCulloch}}, \bibinfo {author} {\bibfnamefont {C.-S.}\
  \bibnamefont {Ha}}, \ and\ \bibinfo {author} {\bibfnamefont {M.}~\bibnamefont
  {Ree}},\ }\href {\doibase 10.1038/nmat1574} {\bibfield  {journal} {\bibinfo
  {journal} {Nature Mater.}\ }\textbf {\bibinfo {volume} {5}},\ \bibinfo
  {pages} {197} (\bibinfo {year} {2006})}\BibitemShut {NoStop}%
\bibitem [{\citenamefont {Deibel}\ \emph {et~al.}(2008)\citenamefont {Deibel},
  \citenamefont {Baumann},\ and\ \citenamefont {Dyakonov}}]{deibel2008}%
  \BibitemOpen
  \bibfield  {author} {\bibinfo {author} {\bibfnamefont {C.}~\bibnamefont
  {Deibel}}, \bibinfo {author} {\bibfnamefont {A.}~\bibnamefont {Baumann}}, \
  and\ \bibinfo {author} {\bibfnamefont {V.}~\bibnamefont {Dyakonov}},\ }\href
  {\doibase 10.1063/1.3005593} {\bibfield  {journal} {\bibinfo  {journal}
  {Appl. Phys. Lett.}\ }\textbf {\bibinfo {volume} {93}},\ \bibinfo {pages}
  {163303} (\bibinfo {year} {2008})}\BibitemShut {NoStop}%
\bibitem [{\citenamefont {Ju\v{s}ka}\ \emph {et~al.}(2008)\citenamefont
  {Ju\v{s}ka}, \citenamefont {Genevi\v{c}ius}, \citenamefont {Nekra\v{s}as},
  \citenamefont {Sliau\v{z}ys},\ and\ \citenamefont {Dennler}}]{juska2008b}%
  \BibitemOpen
  \bibfield  {author} {\bibinfo {author} {\bibfnamefont {G.}~\bibnamefont
  {Ju\v{s}ka}}, \bibinfo {author} {\bibfnamefont {K.}~\bibnamefont
  {Genevi\v{c}ius}}, \bibinfo {author} {\bibfnamefont {N.}~\bibnamefont
  {Nekra\v{s}as}}, \bibinfo {author} {\bibfnamefont {G.}~\bibnamefont
  {Sliau\v{z}ys}}, \ and\ \bibinfo {author} {\bibfnamefont {G.}~\bibnamefont
  {Dennler}},\ }\href {\doibase 10.1063/1.2996588} {\bibfield  {journal}
  {\bibinfo  {journal} {Appl. Phys. Lett.}\ }\textbf {\bibinfo {volume} {93}},\
  \bibinfo {pages} {143303} (\bibinfo {year} {2008})}\BibitemShut {NoStop}%
\bibitem [{\citenamefont {Clarke}\ \emph
  {et~al.}(2012{\natexlab{a}})\citenamefont {Clarke}, \citenamefont {Peet},
  \citenamefont {Denk}, \citenamefont {Dennler}, \citenamefont
  {Lungenschmied},\ and\ \citenamefont {Mozer}}]{clarke2012a}%
  \BibitemOpen
  \bibfield  {author} {\bibinfo {author} {\bibfnamefont {T.~M.}\ \bibnamefont
  {Clarke}}, \bibinfo {author} {\bibfnamefont {J.}~\bibnamefont {Peet}},
  \bibinfo {author} {\bibfnamefont {P.}~\bibnamefont {Denk}}, \bibinfo {author}
  {\bibfnamefont {G.}~\bibnamefont {Dennler}}, \bibinfo {author} {\bibfnamefont
  {C.}~\bibnamefont {Lungenschmied}}, \ and\ \bibinfo {author} {\bibfnamefont
  {A.~J.}\ \bibnamefont {Mozer}},\ }\href {\doibase 10.1039/c1ee02434e}
  {\bibfield  {journal} {\bibinfo  {journal} {Energy Environ. Sci.}\ }\textbf
  {\bibinfo {volume} {5}},\ \bibinfo {pages} {5241} (\bibinfo {year}
  {2012}{\natexlab{a}})}\BibitemShut {NoStop}%
\bibitem [{\citenamefont {Rauh}\ \emph {et~al.}(2012)\citenamefont {Rauh},
  \citenamefont {Deibel},\ and\ \citenamefont {Dyakonov}}]{rauh2012}%
  \BibitemOpen
  \bibfield  {author} {\bibinfo {author} {\bibfnamefont {D.}~\bibnamefont
  {Rauh}}, \bibinfo {author} {\bibfnamefont {C.}~\bibnamefont {Deibel}}, \ and\
  \bibinfo {author} {\bibfnamefont {V.}~\bibnamefont {Dyakonov}},\ }\href
  {\doibase 10.1002/adfm.201103118} {\bibfield  {journal} {\bibinfo  {journal}
  {Adv. Funct. Mater.}\ }\textbf {\bibinfo {volume} {22}},\ \bibinfo {pages}
  {3371} (\bibinfo {year} {2012})}\BibitemShut {NoStop}%
\bibitem [{\citenamefont {Foertig}\ \emph {et~al.}(2013)\citenamefont
  {Foertig}, \citenamefont {Kniepert}, \citenamefont {Gluecker}, \citenamefont
  {Brenner}, \citenamefont {Dyakonov}, \citenamefont {Neher},\ and\
  \citenamefont {Deibel}}]{foertig2013}%
  \BibitemOpen
  \bibfield  {author} {\bibinfo {author} {\bibfnamefont {A.}~\bibnamefont
  {Foertig}}, \bibinfo {author} {\bibfnamefont {J.}~\bibnamefont {Kniepert}},
  \bibinfo {author} {\bibfnamefont {M.}~\bibnamefont {Gluecker}}, \bibinfo
  {author} {\bibfnamefont {T.}~\bibnamefont {Brenner}}, \bibinfo {author}
  {\bibfnamefont {V.}~\bibnamefont {Dyakonov}}, \bibinfo {author}
  {\bibfnamefont {D.}~\bibnamefont {Neher}}, \ and\ \bibinfo {author}
  {\bibfnamefont {C.}~\bibnamefont {Deibel}},\ }\href {\doibase
  10.1002/adfm.201302134} {\bibfield  {journal} {\bibinfo  {journal} {Adv.
  Funct. Mater.}\ }\textbf {\bibinfo {volume} {24}},\ \bibinfo {pages} {1306}
  (\bibinfo {year} {2013})}\BibitemShut {NoStop}%
\bibitem [{\citenamefont {Clarke}\ \emph {et~al.}(2009)\citenamefont {Clarke},
  \citenamefont {Jamieson},\ and\ \citenamefont {Durrant}}]{clarke2009}%
  \BibitemOpen
  \bibfield  {author} {\bibinfo {author} {\bibfnamefont {T.~M.}\ \bibnamefont
  {Clarke}}, \bibinfo {author} {\bibfnamefont {F.~C.}\ \bibnamefont
  {Jamieson}}, \ and\ \bibinfo {author} {\bibfnamefont {J.~R.}\ \bibnamefont
  {Durrant}},\ }\href {\doibase 10.1021/jp909442s} {\bibfield  {journal}
  {\bibinfo  {journal} {J. Phys. Chem. C}\ }\textbf {\bibinfo {volume} {113}},\
  \bibinfo {pages} {20934} (\bibinfo {year} {2009})}\BibitemShut {NoStop}%
\bibitem [{\citenamefont {Foertig}\ \emph {et~al.}(2009)\citenamefont
  {Foertig}, \citenamefont {Baumann}, \citenamefont {Rauh}, \citenamefont
  {Dyakonov},\ and\ \citenamefont {Deibel}}]{foertig2009}%
  \BibitemOpen
  \bibfield  {author} {\bibinfo {author} {\bibfnamefont {A.}~\bibnamefont
  {Foertig}}, \bibinfo {author} {\bibfnamefont {A.}~\bibnamefont {Baumann}},
  \bibinfo {author} {\bibfnamefont {D.}~\bibnamefont {Rauh}}, \bibinfo {author}
  {\bibfnamefont {V.}~\bibnamefont {Dyakonov}}, \ and\ \bibinfo {author}
  {\bibfnamefont {C.}~\bibnamefont {Deibel}},\ }\href {\doibase
  10.1063/1.3202389} {\bibfield  {journal} {\bibinfo  {journal} {Appl. Phys.
  Lett.}\ }\textbf {\bibinfo {volume} {95}},\ \bibinfo {pages} {052104}
  (\bibinfo {year} {2009})}\BibitemShut {NoStop}%
\bibitem [{\citenamefont {Ju\v{s}ka}\ \emph {et~al.}(2009)\citenamefont
  {Ju\v{s}ka}, \citenamefont {Genevi\v{c}ius}, \citenamefont {Nekra\v{s}as},
  \citenamefont {Sliau\v{z}ys},\ and\ \citenamefont
  {{\"O}sterbacka}}]{juska2009}%
  \BibitemOpen
  \bibfield  {author} {\bibinfo {author} {\bibfnamefont {G.}~\bibnamefont
  {Ju\v{s}ka}}, \bibinfo {author} {\bibfnamefont {K.}~\bibnamefont
  {Genevi\v{c}ius}}, \bibinfo {author} {\bibfnamefont {N.}~\bibnamefont
  {Nekra\v{s}as}}, \bibinfo {author} {\bibfnamefont {G.}~\bibnamefont
  {Sliau\v{z}ys}}, \ and\ \bibinfo {author} {\bibfnamefont {R.}~\bibnamefont
  {{\"O}sterbacka}},\ }\href {\doibase 10.1063/1.3141513} {\bibfield  {journal}
  {\bibinfo  {journal} {Appl. Phys. Lett.}\ }\textbf {\bibinfo {volume} {95}},\
  \bibinfo {pages} {013303} (\bibinfo {year} {2009})}\BibitemShut {NoStop}%
\bibitem [{\citenamefont {Shuttle}\ \emph {et~al.}(2010)\citenamefont
  {Shuttle}, \citenamefont {Hamilton}, \citenamefont {Nelson}, \citenamefont
  {O'Regan},\ and\ \citenamefont {Durrant}}]{shuttle2010a}%
  \BibitemOpen
  \bibfield  {author} {\bibinfo {author} {\bibfnamefont {C.~G.}\ \bibnamefont
  {Shuttle}}, \bibinfo {author} {\bibfnamefont {R.}~\bibnamefont {Hamilton}},
  \bibinfo {author} {\bibfnamefont {J.}~\bibnamefont {Nelson}}, \bibinfo
  {author} {\bibfnamefont {B.~C.}\ \bibnamefont {O'Regan}}, \ and\ \bibinfo
  {author} {\bibfnamefont {J.~R.}\ \bibnamefont {Durrant}},\ }\href {\doibase
  10.1002/adfm.200901734} {\bibfield  {journal} {\bibinfo  {journal} {Adv.
  Funct. Mater.}\ }\textbf {\bibinfo {volume} {20}},\ \bibinfo {pages} {698}
  (\bibinfo {year} {2010})}\BibitemShut {NoStop}%
\bibitem [{\citenamefont {Kirchartz}\ and\ \citenamefont
  {Nelson}(2012)}]{kirchartz2012}%
  \BibitemOpen
  \bibfield  {author} {\bibinfo {author} {\bibfnamefont {T.}~\bibnamefont
  {Kirchartz}}\ and\ \bibinfo {author} {\bibfnamefont {J.}~\bibnamefont
  {Nelson}},\ }\href {\doibase 10.1103/PhysRevB.86.165201} {\bibfield
  {journal} {\bibinfo  {journal} {Phys. Rev. B}\ }\textbf {\bibinfo {volume}
  {86}},\ \bibinfo {pages} {165201} (\bibinfo {year} {2012})}\BibitemShut
  {NoStop}%
\bibitem [{\citenamefont {Deibel}\ \emph {et~al.}(2013)\citenamefont {Deibel},
  \citenamefont {Rauh},\ and\ \citenamefont {Foertig}}]{deibel2013}%
  \BibitemOpen
  \bibfield  {author} {\bibinfo {author} {\bibfnamefont {C.}~\bibnamefont
  {Deibel}}, \bibinfo {author} {\bibfnamefont {D.}~\bibnamefont {Rauh}}, \ and\
  \bibinfo {author} {\bibfnamefont {A.}~\bibnamefont {Foertig}},\ }\href
  {\doibase 10.1063/1.4816720} {\bibfield  {journal} {\bibinfo  {journal}
  {Appl. Phys. Lett.}\ }\textbf {\bibinfo {volume} {103}},\ \bibinfo {pages}
  {043307} (\bibinfo {year} {2013})}\BibitemShut {NoStop}%
\bibitem [{\citenamefont {Pivrikas}\ \emph
  {et~al.}(2005{\natexlab{b}})\citenamefont {Pivrikas}, \citenamefont
  {Ju{\v{s}}ka}, \citenamefont {Mozer}, \citenamefont {Scharber}, \citenamefont
  {Arlauskas}, \citenamefont {Sariciftci}, \citenamefont {Stubb},\ and\
  \citenamefont {{\"{O}}sterbacka}}]{pivrikas2005b}%
  \BibitemOpen
  \bibfield  {author} {\bibinfo {author} {\bibfnamefont {A.}~\bibnamefont
  {Pivrikas}}, \bibinfo {author} {\bibfnamefont {G.}~\bibnamefont
  {Ju{\v{s}}ka}}, \bibinfo {author} {\bibfnamefont {A.~J.}\ \bibnamefont
  {Mozer}}, \bibinfo {author} {\bibfnamefont {M.}~\bibnamefont {Scharber}},
  \bibinfo {author} {\bibfnamefont {K.}~\bibnamefont {Arlauskas}}, \bibinfo
  {author} {\bibfnamefont {N.~S.}\ \bibnamefont {Sariciftci}}, \bibinfo
  {author} {\bibfnamefont {H.}~\bibnamefont {Stubb}}, \ and\ \bibinfo {author}
  {\bibfnamefont {R.}~\bibnamefont {{\"{O}}sterbacka}},\ }\href {\doibase
  10.1103/PhysRevLett.94.176806} {\bibfield  {journal} {\bibinfo  {journal}
  {Phys. Rev. Lett.}\ }\textbf {\bibinfo {volume} {94}},\ \bibinfo {pages}
  {176806} (\bibinfo {year} {2005}{\natexlab{b}})}\BibitemShut {NoStop}%
\bibitem [{\citenamefont {Ju\v{s}ka}\ \emph {et~al.}(2005)\citenamefont
  {Ju\v{s}ka}, \citenamefont {Arlauskas}, \citenamefont {Sliau{\v{z}}ys},
  \citenamefont {Pivrikas}, \citenamefont {Mozer}, \citenamefont {Sariciftci},
  \citenamefont {Scharber},\ and\ \citenamefont
  {{\"{O}}sterbacka}}]{juska2005}%
  \BibitemOpen
  \bibfield  {author} {\bibinfo {author} {\bibfnamefont {G.}~\bibnamefont
  {Ju\v{s}ka}}, \bibinfo {author} {\bibfnamefont {K.}~\bibnamefont
  {Arlauskas}}, \bibinfo {author} {\bibfnamefont {G.}~\bibnamefont
  {Sliau{\v{z}}ys}}, \bibinfo {author} {\bibfnamefont {A.}~\bibnamefont
  {Pivrikas}}, \bibinfo {author} {\bibfnamefont {A.~J.}\ \bibnamefont {Mozer}},
  \bibinfo {author} {\bibfnamefont {N.~S.}\ \bibnamefont {Sariciftci}},
  \bibinfo {author} {\bibfnamefont {M.}~\bibnamefont {Scharber}}, \ and\
  \bibinfo {author} {\bibfnamefont {R.}~\bibnamefont {{\"{O}}sterbacka}},\
  }\href {\doibase 10.1063/1.2137454} {\bibfield  {journal} {\bibinfo
  {journal} {Appl. Phys. Lett.}\ }\textbf {\bibinfo {volume} {87}},\ \bibinfo
  {pages} {222110} (\bibinfo {year} {2005})}\BibitemShut {NoStop}%
\bibitem [{\citenamefont {Shuttle}\ \emph {et~al.}(2008)\citenamefont
  {Shuttle}, \citenamefont {O'Regan}, \citenamefont {Ballantyne}, \citenamefont
  {Nelson}, \citenamefont {Bradley},\ and\ \citenamefont
  {Durrant}}]{shuttle2008b}%
  \BibitemOpen
  \bibfield  {author} {\bibinfo {author} {\bibfnamefont {C.~G.}\ \bibnamefont
  {Shuttle}}, \bibinfo {author} {\bibfnamefont {B.}~\bibnamefont {O'Regan}},
  \bibinfo {author} {\bibfnamefont {A.~M.}\ \bibnamefont {Ballantyne}},
  \bibinfo {author} {\bibfnamefont {J.}~\bibnamefont {Nelson}}, \bibinfo
  {author} {\bibfnamefont {D.~D.~C.}\ \bibnamefont {Bradley}}, \ and\ \bibinfo
  {author} {\bibfnamefont {J.~R.}\ \bibnamefont {Durrant}},\ }\href {\doibase
  10.1103/PhysRevB.78.113201} {\bibfield  {journal} {\bibinfo  {journal} {Phys.
  Rev. B}\ }\textbf {\bibinfo {volume} {78}},\ \bibinfo {pages} {113201}
  (\bibinfo {year} {2008})}\BibitemShut {NoStop}%
\bibitem [{\citenamefont {Koster}\ \emph {et~al.}(2006)\citenamefont {Koster},
  \citenamefont {Mihailetchi},\ and\ \citenamefont {Blom}}]{koster2006a}%
  \BibitemOpen
  \bibfield  {author} {\bibinfo {author} {\bibfnamefont {L.~J.~A.}\
  \bibnamefont {Koster}}, \bibinfo {author} {\bibfnamefont {V.~D.}\
  \bibnamefont {Mihailetchi}}, \ and\ \bibinfo {author} {\bibfnamefont
  {P.~W.~M.}\ \bibnamefont {Blom}},\ }\href {\doibase 10.1063/1.2170424}
  {\bibfield  {journal} {\bibinfo  {journal} {Appl. Phys. Lett.}\ }\textbf
  {\bibinfo {volume} {88}},\ \bibinfo {pages} {052104} (\bibinfo {year}
  {2006})}\BibitemShut {NoStop}%
\bibitem [{\citenamefont {Scharber}\ \emph {et~al.}(2010)\citenamefont
  {Scharber}, \citenamefont {Koppe}, , \citenamefont {Gao}, \citenamefont
  {Cordella}, \citenamefont {Loi}, \citenamefont {Denk}, \citenamefont
  {Morana}, \citenamefont {Egelhaaf}, \citenamefont {Forberich}, \citenamefont
  {Dennler}, \citenamefont {Gaudiana}, \citenamefont {Waller}, \citenamefont
  {Zhu}, \citenamefont {Shi},\ and\ \citenamefont {Brabec}}]{scharber2010}%
  \BibitemOpen
  \bibfield  {author} {\bibinfo {author} {\bibfnamefont {M.~C.}\ \bibnamefont
  {Scharber}}, \bibinfo {author} {\bibfnamefont {M.}~\bibnamefont {Koppe}}, ,
  \bibinfo {author} {\bibfnamefont {J.}~\bibnamefont {Gao}}, \bibinfo {author}
  {\bibfnamefont {F.}~\bibnamefont {Cordella}}, \bibinfo {author}
  {\bibfnamefont {M.~A.}\ \bibnamefont {Loi}}, \bibinfo {author} {\bibfnamefont
  {P.}~\bibnamefont {Denk}}, \bibinfo {author} {\bibfnamefont {M.}~\bibnamefont
  {Morana}}, \bibinfo {author} {\bibfnamefont {H.-J.}\ \bibnamefont
  {Egelhaaf}}, \bibinfo {author} {\bibfnamefont {K.}~\bibnamefont {Forberich}},
  \bibinfo {author} {\bibfnamefont {G.}~\bibnamefont {Dennler}}, \bibinfo
  {author} {\bibfnamefont {R.}~\bibnamefont {Gaudiana}}, \bibinfo {author}
  {\bibfnamefont {D.}~\bibnamefont {Waller}}, \bibinfo {author} {\bibfnamefont
  {Z.}~\bibnamefont {Zhu}}, \bibinfo {author} {\bibfnamefont {X.}~\bibnamefont
  {Shi}}, \ and\ \bibinfo {author} {\bibfnamefont {C.~J.}\ \bibnamefont
  {Brabec}},\ }\href {\doibase 10.1002/adma.200900529} {\bibfield  {journal}
  {\bibinfo  {journal} {Adv. Mater.}\ }\textbf {\bibinfo {volume} {22}},\
  \bibinfo {pages} {367} (\bibinfo {year} {2010})}\BibitemShut {NoStop}%
\bibitem [{\citenamefont {Clarke}\ \emph {et~al.}(2011)\citenamefont {Clarke},
  \citenamefont {Rodovsky}, \citenamefont {Herzing}, \citenamefont {Peet},
  \citenamefont {Dennler}, \citenamefont {DeLongchamp}, \citenamefont
  {Lungenschmied},\ and\ \citenamefont {Mozer}}]{clarke2011}%
  \BibitemOpen
  \bibfield  {author} {\bibinfo {author} {\bibfnamefont {T.~M.}\ \bibnamefont
  {Clarke}}, \bibinfo {author} {\bibfnamefont {D.~B.}\ \bibnamefont
  {Rodovsky}}, \bibinfo {author} {\bibfnamefont {A.~A.}\ \bibnamefont
  {Herzing}}, \bibinfo {author} {\bibfnamefont {J.}~\bibnamefont {Peet}},
  \bibinfo {author} {\bibfnamefont {G.}~\bibnamefont {Dennler}}, \bibinfo
  {author} {\bibfnamefont {D.}~\bibnamefont {DeLongchamp}}, \bibinfo {author}
  {\bibfnamefont {C.}~\bibnamefont {Lungenschmied}}, \ and\ \bibinfo {author}
  {\bibfnamefont {A.~J.}\ \bibnamefont {Mozer}},\ }\href {\doibase
  10.1002/aenm.201100390} {\bibfield  {journal} {\bibinfo  {journal} {Adv.
  Energy Mater.}\ }\textbf {\bibinfo {volume} {1}},\ \bibinfo {pages} {1062}
  (\bibinfo {year} {2011})}\BibitemShut {NoStop}%
\bibitem [{\citenamefont {Murthy}\ \emph {et~al.}(2013)\citenamefont {Murthy},
  \citenamefont {Melianas}, \citenamefont {Tang}, \citenamefont {Ju\v{s}ka},
  \citenamefont {Arlauskas}, \citenamefont {Zhang}, \citenamefont {Siebbeles},
  \citenamefont {Ingan{\"a}s},\ and\ \citenamefont {Savenije}}]{murthy2013}%
  \BibitemOpen
  \bibfield  {author} {\bibinfo {author} {\bibfnamefont {D.~H.~K.}\
  \bibnamefont {Murthy}}, \bibinfo {author} {\bibfnamefont {A.}~\bibnamefont
  {Melianas}}, \bibinfo {author} {\bibfnamefont {Z.}~\bibnamefont {Tang}},
  \bibinfo {author} {\bibfnamefont {G.}~\bibnamefont {Ju\v{s}ka}}, \bibinfo
  {author} {\bibfnamefont {K.}~\bibnamefont {Arlauskas}}, \bibinfo {author}
  {\bibfnamefont {F.}~\bibnamefont {Zhang}}, \bibinfo {author} {\bibfnamefont
  {L.~D.~A.}\ \bibnamefont {Siebbeles}}, \bibinfo {author} {\bibfnamefont
  {O.}~\bibnamefont {Ingan{\"a}s}}, \ and\ \bibinfo {author} {\bibfnamefont
  {T.~J.}\ \bibnamefont {Savenije}},\ }\href {\doibase 10.1002/adfm.201203852}
  {\bibfield  {journal} {\bibinfo  {journal} {Adv. Funct. Mater.}\ }\textbf
  {\bibinfo {volume} {23}},\ \bibinfo {pages} {4262} (\bibinfo {year}
  {2013})}\BibitemShut {NoStop}%
\bibitem [{\citenamefont {Albrecht}\ \emph {et~al.}(2012)\citenamefont
  {Albrecht}, \citenamefont {Janietz}, \citenamefont {Schindler}, \citenamefont
  {Frisch}, \citenamefont {Kurpiers}, \citenamefont {Kniepert}, \citenamefont
  {Inal}, \citenamefont {Pingel}, \citenamefont {Fostiropoulos}, \citenamefont
  {Koch},\ and\ \citenamefont {Neher}}]{albrecht2012}%
  \BibitemOpen
  \bibfield  {author} {\bibinfo {author} {\bibfnamefont {S.}~\bibnamefont
  {Albrecht}}, \bibinfo {author} {\bibfnamefont {S.}~\bibnamefont {Janietz}},
  \bibinfo {author} {\bibfnamefont {W.}~\bibnamefont {Schindler}}, \bibinfo
  {author} {\bibfnamefont {J.}~\bibnamefont {Frisch}}, \bibinfo {author}
  {\bibfnamefont {J.}~\bibnamefont {Kurpiers}}, \bibinfo {author}
  {\bibfnamefont {J.}~\bibnamefont {Kniepert}}, \bibinfo {author}
  {\bibfnamefont {S.}~\bibnamefont {Inal}}, \bibinfo {author} {\bibfnamefont
  {P.}~\bibnamefont {Pingel}}, \bibinfo {author} {\bibfnamefont
  {K.}~\bibnamefont {Fostiropoulos}}, \bibinfo {author} {\bibfnamefont
  {N.}~\bibnamefont {Koch}}, \ and\ \bibinfo {author} {\bibfnamefont
  {D.}~\bibnamefont {Neher}},\ }\href {\doibase 10.1021/ja305039j} {\bibfield
  {journal} {\bibinfo  {journal} {J. Am. Chem. Soc.}\ }\textbf {\bibinfo
  {volume} {134}},\ \bibinfo {pages} {14932} (\bibinfo {year}
  {2012})}\BibitemShut {NoStop}%
\bibitem [{\citenamefont {Wetzelaer}\ \emph {et~al.}(2013)\citenamefont
  {Wetzelaer}, \citenamefont {van~der Kaap}, \citenamefont {Koster},\ and\
  \citenamefont {Blom}}]{wetzelaer2013}%
  \BibitemOpen
  \bibfield  {author} {\bibinfo {author} {\bibfnamefont {G.-J. A.~H.}\
  \bibnamefont {Wetzelaer}}, \bibinfo {author} {\bibfnamefont {N.~J.}\
  \bibnamefont {van~der Kaap}}, \bibinfo {author} {\bibfnamefont {L.~J.~A.}\
  \bibnamefont {Koster}}, \ and\ \bibinfo {author} {\bibfnamefont {P.~W.~M.}\
  \bibnamefont {Blom}},\ }\href {\doibase 10.1002/aenm.201300251} {\bibfield
  {journal} {\bibinfo  {journal} {Adv. Energy Mater.}\ }\textbf {\bibinfo
  {volume} {3}},\ \bibinfo {pages} {1130} (\bibinfo {year} {2013})}\BibitemShut
  {NoStop}%
\bibitem [{\citenamefont {Roland}\ \emph {et~al.}(2014)\citenamefont {Roland},
  \citenamefont {Schubert}, \citenamefont {Collins}, \citenamefont {Kurpiers},
  \citenamefont {Chen}, \citenamefont {Facchetti}, \citenamefont {Ade},\ and\
  \citenamefont {Neher}}]{roland2014}%
  \BibitemOpen
  \bibfield  {author} {\bibinfo {author} {\bibfnamefont {S.}~\bibnamefont
  {Roland}}, \bibinfo {author} {\bibfnamefont {M.}~\bibnamefont {Schubert}},
  \bibinfo {author} {\bibfnamefont {B.~A.}\ \bibnamefont {Collins}}, \bibinfo
  {author} {\bibfnamefont {J.}~\bibnamefont {Kurpiers}}, \bibinfo {author}
  {\bibfnamefont {Z.}~\bibnamefont {Chen}}, \bibinfo {author} {\bibfnamefont
  {A.}~\bibnamefont {Facchetti}}, \bibinfo {author} {\bibfnamefont
  {H.}~\bibnamefont {Ade}}, \ and\ \bibinfo {author} {\bibfnamefont
  {D.}~\bibnamefont {Neher}},\ }\href {\doibase 10.1021/jz501506z} {\bibfield
  {journal} {\bibinfo  {journal} {J. Phys. Chem. Lett.}\ }\textbf {\bibinfo
  {volume} {5}},\ \bibinfo {pages} {2815} (\bibinfo {year} {2014})}\BibitemShut
  {NoStop}%
\bibitem [{\citenamefont {Mozer}\ \emph {et~al.}(2005)\citenamefont {Mozer},
  \citenamefont {Dennler}, \citenamefont {Sariciftci}, \citenamefont
  {Westerling}, \citenamefont {Pivrikas}, \citenamefont {{\"{O}}sterbacka},\
  and\ \citenamefont {Ju{\v{s}}ka}}]{mozer2005c}%
  \BibitemOpen
  \bibfield  {author} {\bibinfo {author} {\bibfnamefont {A.~J.}\ \bibnamefont
  {Mozer}}, \bibinfo {author} {\bibfnamefont {G.}~\bibnamefont {Dennler}},
  \bibinfo {author} {\bibfnamefont {N.~S.}\ \bibnamefont {Sariciftci}},
  \bibinfo {author} {\bibfnamefont {M.}~\bibnamefont {Westerling}}, \bibinfo
  {author} {\bibfnamefont {A.}~\bibnamefont {Pivrikas}}, \bibinfo {author}
  {\bibfnamefont {R.}~\bibnamefont {{\"{O}}sterbacka}}, \ and\ \bibinfo
  {author} {\bibfnamefont {G.}~\bibnamefont {Ju{\v{s}}ka}},\ }\href {\doibase
  10.1103/PhysRevB.72.035217} {\bibfield  {journal} {\bibinfo  {journal} {Phys.
  Rev. B}\ }\textbf {\bibinfo {volume} {72}},\ \bibinfo {pages} {035217}
  (\bibinfo {year} {2005})}\BibitemShut {NoStop}%
\bibitem [{\citenamefont {Dennler}\ \emph {et~al.}(2006)\citenamefont
  {Dennler}, \citenamefont {Mozer}, \citenamefont {Ju{\v{s}}ka}, \citenamefont
  {Pivrikas}, \citenamefont {{\"{O}}sterbacka}, \citenamefont {Fuchsbauer},\
  and\ \citenamefont {Sariciftci}}]{dennler2006}%
  \BibitemOpen
  \bibfield  {author} {\bibinfo {author} {\bibfnamefont {G.}~\bibnamefont
  {Dennler}}, \bibinfo {author} {\bibfnamefont {A.~J.}\ \bibnamefont {Mozer}},
  \bibinfo {author} {\bibfnamefont {G.}~\bibnamefont {Ju{\v{s}}ka}}, \bibinfo
  {author} {\bibfnamefont {A.}~\bibnamefont {Pivrikas}}, \bibinfo {author}
  {\bibfnamefont {R.}~\bibnamefont {{\"{O}}sterbacka}}, \bibinfo {author}
  {\bibfnamefont {A.}~\bibnamefont {Fuchsbauer}}, \ and\ \bibinfo {author}
  {\bibfnamefont {N.~S.}\ \bibnamefont {Sariciftci}},\ }\href {\doibase
  10.1016/j.orgel.2006.02.004} {\bibfield  {journal} {\bibinfo  {journal} {Org.
  Electron.}\ }\textbf {\bibinfo {volume} {7}},\ \bibinfo {pages} {229}
  (\bibinfo {year} {2006})}\BibitemShut {NoStop}%
\bibitem [{\citenamefont {Clarke}\ \emph
  {et~al.}(2012{\natexlab{b}})\citenamefont {Clarke}, \citenamefont {Peet},
  \citenamefont {Nattestad}, \citenamefont {Drolet}, \citenamefont {Dennler},
  \citenamefont {Lungenschmied}, \citenamefont {Leclerc},\ and\ \citenamefont
  {Mozer}}]{clarke2012b}%
  \BibitemOpen
  \bibfield  {author} {\bibinfo {author} {\bibfnamefont {T.~M.}\ \bibnamefont
  {Clarke}}, \bibinfo {author} {\bibfnamefont {J.}~\bibnamefont {Peet}},
  \bibinfo {author} {\bibfnamefont {A.}~\bibnamefont {Nattestad}}, \bibinfo
  {author} {\bibfnamefont {N.}~\bibnamefont {Drolet}}, \bibinfo {author}
  {\bibfnamefont {G.}~\bibnamefont {Dennler}}, \bibinfo {author} {\bibfnamefont
  {C.}~\bibnamefont {Lungenschmied}}, \bibinfo {author} {\bibfnamefont
  {M.}~\bibnamefont {Leclerc}}, \ and\ \bibinfo {author} {\bibfnamefont
  {A.~J.}\ \bibnamefont {Mozer}},\ }\href {\doibase
  10.1016/j.orgel.2012.07.037} {\bibfield  {journal} {\bibinfo  {journal} {Org.
  Electron.}\ }\textbf {\bibinfo {volume} {13}},\ \bibinfo {pages} {2639}
  (\bibinfo {year} {2012}{\natexlab{b}})}\BibitemShut {NoStop}%
\bibitem [{\citenamefont {Szmytkowski}(2009)}]{szmytkowski2009}%
  \BibitemOpen
  \bibfield  {author} {\bibinfo {author} {\bibfnamefont {J.}~\bibnamefont
  {Szmytkowski}},\ }\href {\doibase 10.1016/j.cplett.2009.01.043} {\bibfield
  {journal} {\bibinfo  {journal} {Chem. Phys. Lett.}\ }\textbf {\bibinfo
  {volume} {470}},\ \bibinfo {pages} {123} (\bibinfo {year}
  {2009})}\BibitemShut {NoStop}%
\bibitem [{\citenamefont {Deibel}\ \emph {et~al.}(2009)\citenamefont {Deibel},
  \citenamefont {Wagenpfahl},\ and\ \citenamefont {Dyakonov}}]{deibel2009b}%
  \BibitemOpen
  \bibfield  {author} {\bibinfo {author} {\bibfnamefont {C.}~\bibnamefont
  {Deibel}}, \bibinfo {author} {\bibfnamefont {A.}~\bibnamefont {Wagenpfahl}},
  \ and\ \bibinfo {author} {\bibfnamefont {V.}~\bibnamefont {Dyakonov}},\
  }\href {\doibase 10.1103/PhysRevB.80.075203} {\bibfield  {journal} {\bibinfo
  {journal} {Phys. Rev. B}\ }\textbf {\bibinfo {volume} {80}},\ \bibinfo
  {pages} {075203} (\bibinfo {year} {2009})}\BibitemShut {NoStop}%
\bibitem [{\citenamefont {Hilczer}\ and\ \citenamefont
  {Tachiya}(2010)}]{hilczer2010}%
  \BibitemOpen
  \bibfield  {author} {\bibinfo {author} {\bibfnamefont {M.}~\bibnamefont
  {Hilczer}}\ and\ \bibinfo {author} {\bibfnamefont {M.}~\bibnamefont
  {Tachiya}},\ }\href {\doibase 10.1021/jp912262h} {\bibfield  {journal}
  {\bibinfo  {journal} {J. Phys. Chem.}\ }\textbf {\bibinfo {volume} {114}},\
  \bibinfo {pages} {6808} (\bibinfo {year} {2010})}\BibitemShut {NoStop}%
\bibitem [{\citenamefont {Howard}\ \emph {et~al.}(2010)\citenamefont {Howard},
  \citenamefont {Mauer}, \citenamefont {Meister},\ and\ \citenamefont
  {Laquai}}]{howard2010b}%
  \BibitemOpen
  \bibfield  {author} {\bibinfo {author} {\bibfnamefont {I.~A.}\ \bibnamefont
  {Howard}}, \bibinfo {author} {\bibfnamefont {R.}~\bibnamefont {Mauer}},
  \bibinfo {author} {\bibfnamefont {M.}~\bibnamefont {Meister}}, \ and\
  \bibinfo {author} {\bibfnamefont {F.}~\bibnamefont {Laquai}},\ }\href
  {\doibase 10.1021/ja105260d} {\bibfield  {journal} {\bibinfo  {journal} {J.
  Am. Chem. Soc.}\ }\textbf {\bibinfo {volume} {132}},\ \bibinfo {pages}
  {14866} (\bibinfo {year} {2010})}\BibitemShut {NoStop}%
\bibitem [{\citenamefont {Ferguson}\ \emph {et~al.}(2011)\citenamefont
  {Ferguson}, \citenamefont {Kopidakis}, \citenamefont {Shaheen},\ and\
  \citenamefont {Rumbles}}]{ferguson2011}%
  \BibitemOpen
  \bibfield  {author} {\bibinfo {author} {\bibfnamefont {A.~J.}\ \bibnamefont
  {Ferguson}}, \bibinfo {author} {\bibfnamefont {N.}~\bibnamefont {Kopidakis}},
  \bibinfo {author} {\bibfnamefont {S.~E.}\ \bibnamefont {Shaheen}}, \ and\
  \bibinfo {author} {\bibfnamefont {G.}~\bibnamefont {Rumbles}},\ }\href
  {\doibase 10.1021/jp208014v} {\bibfield  {journal} {\bibinfo  {journal} {J.
  Phys. Chem. C}\ }\textbf {\bibinfo {volume} {115}},\ \bibinfo {pages} {23134}
  (\bibinfo {year} {2011})}\BibitemShut {NoStop}%
\bibitem [{\citenamefont {Hamilton}\ \emph {et~al.}(2010)\citenamefont
  {Hamilton}, \citenamefont {Shuttle}, \citenamefont {O'Regan}, \citenamefont
  {Hammant}, \citenamefont {Nelson},\ and\ \citenamefont
  {Durrant}}]{hamilton2010}%
  \BibitemOpen
  \bibfield  {author} {\bibinfo {author} {\bibfnamefont {R.}~\bibnamefont
  {Hamilton}}, \bibinfo {author} {\bibfnamefont {C.~G.}\ \bibnamefont
  {Shuttle}}, \bibinfo {author} {\bibfnamefont {B.}~\bibnamefont {O'Regan}},
  \bibinfo {author} {\bibfnamefont {T.~C.}\ \bibnamefont {Hammant}}, \bibinfo
  {author} {\bibfnamefont {J.}~\bibnamefont {Nelson}}, \ and\ \bibinfo {author}
  {\bibfnamefont {J.~R.}\ \bibnamefont {Durrant}},\ }\href {\doibase
  10.1021/jz1001506} {\bibfield  {journal} {\bibinfo  {journal} {J. Phys. Chem.
  Lett.}\ }\textbf {\bibinfo {volume} {1}},\ \bibinfo {pages} {1432} (\bibinfo
  {year} {2010})}\BibitemShut {NoStop}%
\bibitem [{\citenamefont {Heiber}(2014)}]{heiber2014a}%
  \BibitemOpen
  \bibfield  {author} {\bibinfo {author} {\bibfnamefont {M.~C.}\ \bibnamefont
  {Heiber}},\ }\href@noop {} {\enquote {\bibinfo {title} {{I}sing\_{OPV}},}\
  }\bibinfo {howpublished} {\url{https://github.com/MikeHeiber/Ising_OPV}}
  (\bibinfo {year} {2014})\BibitemShut {NoStop}%
\bibitem [{\citenamefont {Heiber}\ and\ \citenamefont
  {Dhinojwala}(2014)}]{heiber2014c}%
  \BibitemOpen
  \bibfield  {author} {\bibinfo {author} {\bibfnamefont {M.~C.}\ \bibnamefont
  {Heiber}}\ and\ \bibinfo {author} {\bibfnamefont {A.}~\bibnamefont
  {Dhinojwala}},\ }\href {\doibase 10.1103/PhysRevApplied.2.014008} {\bibfield
  {journal} {\bibinfo  {journal} {Phys. Rev. Appl.}\ }\textbf {\bibinfo
  {volume} {2}},\ \bibinfo {pages} {014008} (\bibinfo {year}
  {2014})}\BibitemShut {NoStop}%
\bibitem [{\citenamefont {Miller}\ and\ \citenamefont
  {Abrahams}(1960)}]{miller1960}%
  \BibitemOpen
  \bibfield  {author} {\bibinfo {author} {\bibfnamefont {A.}~\bibnamefont
  {Miller}}\ and\ \bibinfo {author} {\bibfnamefont {E.}~\bibnamefont
  {Abrahams}},\ }\href {\doibase 10.1103/PhysRev.120.745} {\bibfield  {journal}
  {\bibinfo  {journal} {Phys. Rev.}\ }\textbf {\bibinfo {volume} {120}},\
  \bibinfo {pages} {745} (\bibinfo {year} {1960})}\BibitemShut {NoStop}%
\bibitem [{\citenamefont {Nenashev}\ \emph {et~al.}(2010)\citenamefont
  {Nenashev}, \citenamefont {Jansson}, \citenamefont {Baranovskii},
  \citenamefont {{\"O}sterbacka}, \citenamefont {Dvurechenskii},\ and\
  \citenamefont {Gebhard}}]{nenashev2010b}%
  \BibitemOpen
  \bibfield  {author} {\bibinfo {author} {\bibfnamefont {A.~V.}\ \bibnamefont
  {Nenashev}}, \bibinfo {author} {\bibfnamefont {F.}~\bibnamefont {Jansson}},
  \bibinfo {author} {\bibfnamefont {S.~D.}\ \bibnamefont {Baranovskii}},
  \bibinfo {author} {\bibfnamefont {R.}~\bibnamefont {{\"O}sterbacka}},
  \bibinfo {author} {\bibfnamefont {A.~V.}\ \bibnamefont {Dvurechenskii}}, \
  and\ \bibinfo {author} {\bibfnamefont {F.}~\bibnamefont {Gebhard}},\ }\href
  {\doibase 10.1103/PhysRevB.81.115204} {\bibfield  {journal} {\bibinfo
  {journal} {Phys. Rev. B}\ }\textbf {\bibinfo {volume} {81}},\ \bibinfo
  {pages} {115204} (\bibinfo {year} {2010})}\BibitemShut {NoStop}%
\bibitem [{\citenamefont {Wetzelaer}\ \emph {et~al.}(2011)\citenamefont
  {Wetzelaer}, \citenamefont {Kuik}, \citenamefont {Nicolai},\ and\
  \citenamefont {Blom}}]{wetzelaer2011b}%
  \BibitemOpen
  \bibfield  {author} {\bibinfo {author} {\bibfnamefont {G.~A.~H.}\
  \bibnamefont {Wetzelaer}}, \bibinfo {author} {\bibfnamefont {M.}~\bibnamefont
  {Kuik}}, \bibinfo {author} {\bibfnamefont {H.~T.}\ \bibnamefont {Nicolai}}, \
  and\ \bibinfo {author} {\bibfnamefont {P.~W.~M.}\ \bibnamefont {Blom}},\
  }\href {\doibase 10.1103/PhysRevB.83.165204} {\bibfield  {journal} {\bibinfo
  {journal} {Phys. Rev. B}\ }\textbf {\bibinfo {volume} {83}},\ \bibinfo
  {pages} {165204} (\bibinfo {year} {2011})}\BibitemShut {NoStop}%
\bibitem [{Note1()}]{Note1}%
  \BibitemOpen
  \bibinfo {note} {See Supplemental Material at [URL] for morphology generation
  details, a full list of simulation parameters, and supplementary
  results.}\BibitemShut {Stop}%
\end{thebibliography}

%


\end{document}


\title{Supplementary Information for Encounter-Limited Charge Carrier Recombination in Phase Separated Organic Semiconductor Blends}

\author{Michael C. Heiber}
\email{heiber@mailaps.org}
\affiliation{Experimental Physics VI, Julius-Maximilians-University of W{\"u}rzburg, 97074 W{\"u}rzburg, Germany}
\affiliation{Institut f{\"u}r Physik, Technische Universit{\"a}t Chemnitz, 09126 Chemnitz, Germany}

\author{Christoph Baumbach}
\affiliation{Institut f{\"u}r Physik, Technische Universit{\"a}t Chemnitz, 09126 Chemnitz, Germany}

\author{Vladimir Dyakonov}
\affiliation{Experimental Physics VI, Julius-Maximilians-University of W{\"u}rzburg, 97074 W{\"u}rzburg, Germany}%
\affiliation{Bavarian Centre for Applied Energy Research (ZAE Bayern), 97074 W{\"u}rzburg, Germany}

\author{Carsten Deibel}
\email{deibel@physik.tu-chemnitz.de}
\affiliation{Institut f{\"u}r Physik, Technische Universit{\"a}t Chemnitz, 09126 Chemnitz, Germany}

\date{\today}

\begin{abstract}
\end{abstract}

\pacs{}


\maketitle

\section{Morphology Details}

Morphologies were created using the Ising\_OPV software tool \cite{heiber2014a,heiber2014c} with a 50:50 blend ratio and an interaction energy of $0.6kT$. Utilizing the smoothing and rescaling methods,\cite{heiber2014c} morphologies were generated on lattices with a final size of 100 by 100 by 100 or larger to prevent lattice confinement effects.\cite{heiber2014c} For each set of input parameters, 100 morphologies were independently generated to form a morphology set. Eight morphology sets (MS1,MS2,MS3,MS4,MS5,MS6,MS7,MS8) were generated with different domain sizes of approximately 5, 10, 15, 20, 25, 35, 45 and 55~nm. More detailed information on the input parameters and characterization of each morphology set is provided in Table~\ref{tab:morphology}.

\begin{table*}[ht]
\caption{Morphology Set Information}
\label{tab:morphology}
\begin{tabular}{ l | c | c | c | c | c  | c | c | c}
\hline \hline
& MS1 & MS2 & MS3 & MS4 & MS5 & MS6 & MS7 & MS8 \\
\hline
Initial lattice dimensions & 100 & 50 & 34 & 25 & 23 & 23 & 41 & 28  \\
Monte Carlo steps & 369 & 374 & 374 & 369 & 374 & 369 & 6000 & 1110 \\
Rescale factor & N/A & 2 & 3 & 4 & 5 & 7 & 5 & 3,3\\
Smoothing threshold & 0.52 & 0.52 & 0.52 & 0.52 & 0.52 & 0.52 & 0.52 & 0.52 \\
\hline
Final lattice dimensions & 100 & 100 & 102 & 100 & 115 & 161 & 205 & 252 \\
Domain size, $d$ & 4.92$\pm$0.02 & 10.1$\pm$0.1 & 15.1$\pm$0.3 & 20$\pm$0.7 & 25$\pm$1 & 35$\pm$1 & 45$\pm$2 & 55$\pm$3 \\
Interfacial area/volume & 0.356$\pm$0.001 & 0.175$\pm$0.001 & 0.116$\pm$0.002 & 0.087$\pm$0.002 & 0.069$\pm$0.002 & 0.050$\pm$0.001 & 0.037$\pm$0.001 & 0.031$\pm$0.001 \\
Tortuosity & 1.1$\pm$0.02 & 1.1$\pm$0.03 & 1.1$\pm$0.04 & 1.1$\pm$0.05 & 1.1$\pm$0.05 & 1.1$\pm$0.05 & 1.1$\pm$0.06 & 1.1$\pm$0.06 \\
\hline \hline
\end{tabular}
\end{table*}

\section{KMC Simulation Details}

The KMC simulation methodology used in this study was explained in more detail in a previous paper,\cite{heiber2012} but a short summary of the most important aspects are provided here. The model morphologies were used to define the donor and acceptor sites on a three-dimensional lattice with a lattice constant ($a$) of 1~nm. Both phases were assigned an uncorrelated Gaussian DOS characterized by the energetic disorder parameter ($\sigma$). Periodic boundary conditions were used in two directions to simulate a thin film. To start the simulation, excitons were created with uniform probability throughout the lattice with a Gaussian excitation pulse having a pulse width of 100~ps and an intensity corresponding to an initial exciton concentration of $5 \times 10^{17}$~cm$^{-3}$. 

Exciton diffusion was implemented using the F\"orster resonance energy transfer model,
\begin{equation}
R_{ij,\text{exh}} = R_{0,\text{exh}}\left(\frac{a}{d_{ij}}\right)^6 f_\text{B}(\Delta E_{ij,\text{exh}}),
\end{equation}
where $R_{0,\text{exh}}$ is the exciton hopping prefactor, $d_{ij}$ is the distance between sites, 
\begin{equation}
\label{eq:boltzmann}
f_\text{B}(\Delta E_{ij}) = \begin{cases}  \exp{\left(\frac{-\Delta E_{ij}}{kT} \right)} & \Delta E_{ij}> 0 \\
1 & \Delta E_{ij}\le 0 \end{cases},
\end{equation}
and $\Delta E_{ij,\text{exh}}$ is the change in potential energy for exciton hopping,
\begin{equation}
\Delta E_{ij,\text{exh}} = E_{j,\text{singlet}}-E_{i,\text{singlet}}.
\end{equation}
Exciton hopping events were calculated to sites up to 4~nm away from the starting site. In additon, the exciton relaxation time defines the lifetime of the excited state and is used to calculate the exciton relaxation rate,
\begin{equation}
R_\text{exr} = 1/\tau_\text{ex},
\end{equation}
where $\tau_{ex}$ is the exciton lifetime.

The complexities of charge separation were bypassed to create free charge carriers directly from excitons. To do this, electron-hole pairs were created across the interface with a separation distance of 30~nm by restricting exciton creation to within 30~nm of an interface and executing an ultrafast long-range charge transfer event. Long range charge transfer (exciton dissociation) was implemented using the simplified Miller-Abrahams model where charge transfer is always energetically favorable,
\begin{equation}
R_{ij,\text{exd}} = R_{0,\text{exd}} \exp{(-2 \gamma_\text{ex} d_{ij})}
\end{equation}
where $R_{0,\text{exd}}$ is the exciton dissociation prefactor and $\gamma_{ex}$ is the inverse exciton localization parameter.  Exciton dissociation events were only calculated for sites between 30 and 31~nm away from the starting site.

Charge motion was simulated using the Miller-Abrahams model.  For electrons, 
\begin{equation}
R_\text{ij,\text{elh}} = R_\text{0,\text{e}} \exp({-2 \gamma_\text{ch} d_{ij}}) f_\text{B}(\Delta E_{ij,\text{elh}})
\end{equation}
where $R_\text{0,\text{e}}$ is the electron hopping prefactor, $\gamma_\text{ch}$ is the charge localization parameter, and
\begin{equation}
\Delta E_{ij,\text{elh}} = E_{i,\text{LUMO}}-E_{j,\text{LUMO}}+ \Delta E_{C,ij}-Fd_{ij},
\end{equation}
where $E_{i,\text{LUMO}}$ and $E_{j,\text{LUMO}}$ are the initial and final site energies drawn from the density of states distribution, $\Delta E_{C,ij}$ is the change in Coulomb potential that would occur for hopping from site $i$ to site $j$, and $F$ is the electric field. Analogous expresions are used to calculate the hole hopping rate. Coulomb interactions were included between charges within a cutoff radius.  The cutoff radius should be set to a large enough value such that it does not affect the results of the simulation.  A cutoff radius of 35~nm was found to be large enough to not impact the recombination rate at a charge carrier concentration of 10$^{16}$ cm$^{-3}$. The change in Coulomb potential is calculated,
\begin{equation} \Delta E_{C,ij} = E_{C,j}-E_{C,i}, \end{equation}
where
\begin{equation} E_{C,i} = \sum_{k=1, k\neq i}^N \frac {q_i q_k}{4 \pi \epsilon \epsilon_0 d_{ik}} \qquad d_{ik} \leq35\textrm{ nm}, \end{equation}
given $N$ nearby electrons and holes.  An analogous expression is used to calculate the Coulomb potential for the final state by assuming that the charge of interest is positioned on site j. Electron hopping was restricted to acceptor domains and hole hopping was restricted to donor domains. Charge hopping events were calculated for sites up to 3~nm away from the starting site. 

When an electron and a hole come close together, the charge recombination event is enabled. Charge recombination was also implemented using the Miller-Abrahams model similar to charge hopping,
\begin{equation}
R_{ij,\text{rec}} = R_{0,\text{rec}} \exp{(-2 \gamma_{ch} d_{ij})},
\end{equation}
where $R_{0,rec}$ is the recombination prefactor, which was held constant at a large value of $10^{15}$~s$^{-1}$ to ensure that recombination dominates over redissociation.  Charge recombination events were also calculated for sites up to 3~nm away from the starting site.  In addition, the selective recalculation method \cite{heiber2012} was used with a recalculation cutoff radius of 5~nm. A full list of parameters is provided in Table~\ref{tab:KMC_params}.

\begin{table}[hb]
\begin{ruledtabular}
\caption{KMC Simulation Parameters}
\label{tab:KMC_params}
\begin{tabular}{ll}
Lattice constant, $a$ & 1 nm \\
Temperature, $T$ & 300 K \\
Dielectric constant, $\epsilon$ & 3.5 \\
Energetic disorder, $\sigma$ & 0.075 eV \\
Exciton lifetime, $\tau_{ex}$ & 500 ps \\
Exciton hopping prefactor, $R_{0,\text{exh}}$ & $10^{12}$ s$^{-1}$ \\
Exciton localization, $\gamma_\text{ex}$ & 0.1 nm$^{-1}$ \\
Exciton dissociation prefactor, $R_{0,\text{exd}}$ & 10$^{16}$ s$^{-1}$ \\
Electron hopping prefactor, $R_{0,\text{e}}$ &  10$^{13}$ s$^{-1}$ \\
Hole hopping prefactor, $R_{0,\text{h}}$ & variable \\
Charge localization, $\gamma_\text{ch}$ & 2 nm$^{-1}$ \\
Charge recombination prefactor, $R_{0,\text{rec}}$ & 10$^{15}$ s$^{-1}$ \\
Electric field, $F$ & 0 Vm$^{-1}$\\
Coulomb cutoff radius, $R_\text{cutoff}$ & 35 nm \\
\end{tabular}
\end{ruledtabular}
\end{table}

\section{Supplementary Results}
\begin{figure}
\includegraphics[scale=1]{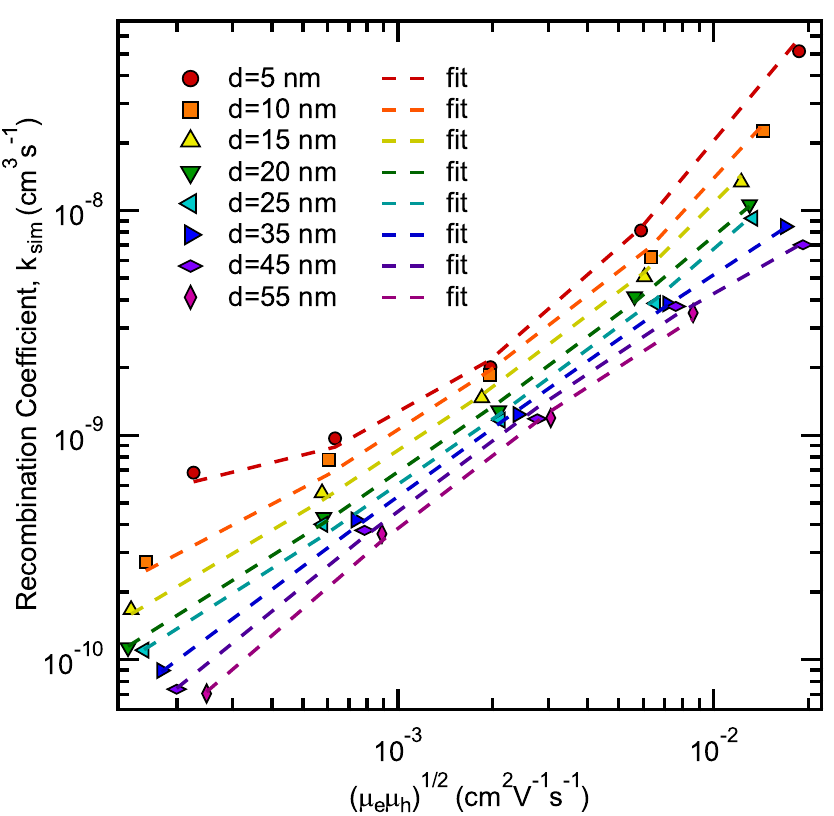}
\caption{\label{fig:S_fits}Fitted recombination coefficient data.}
\end{figure}

All simulated recombination coefficient data was fit with the equation,
\begin{equation}
k_\text{sim} =  \frac{e}{\epsilon\epsilon_0} f_1(d) 2 M_{g(d)}(\mu_e,\mu_h).
\label{eqn:fit}
\end{equation}
where $f_1(d)$ is a domain size dependent prefactor and $M_{g(d)}(\mu_e,\mu_h)$ is the power mean (generalized mean),
\begin{equation}
M_g(\mu_e,\mu_h) = \left(\frac{\mu_e^g+\mu_h^g}{2}\right)^{1/g},
\end{equation}
with a domain size dependent exponent, $g(d)$.
Fig.\ \ref{fig:S_fits} shows the resulting fits of each data set.  The resulting fitted parameters $f_1$ and $g$ and their uncertainties are shown in the main article.  These results show that the power mean in Eqn.\ \ref{eqn:fit} is a very accurate description of the mobility dependence obtained in the recombination simulations.





\bibliography{references}
